\documentclass[sigconf,nonacm]{acmart}
\usepackage{xcolor}
\usepackage{cleveref}
\Crefname{figure}{Fig.}{Figs.}
\usepackage{xspace}
\usepackage{makecell}
\usepackage{csquotes}
\renewcommand{\blockquote}[1]{``\textit{#1}''}
\usepackage{tablefootnote}
\usepackage{booktabs}
\usepackage{enumitem}
\usepackage{threeparttable}
\usepackage{multirow}
\usepackage{varwidth}
\usepackage[most]{tcolorbox}
\usepackage{soul}

\AtBeginDocument{%
  }

\setcopyright{acmlicensed}
\copyrightyear{2018}
\acmYear{2018}
\acmDOI{XXXXXXX.XXXXXXX}

\acmConference[Conference acronym 'XX]{Make sure to enter the correct
  conference title from your rights confirmation emai}{June 03--05,
  2018}{Woodstock, NY}
\acmISBN{978-1-4503-XXXX-X/18/06}

\begin{document}
\newcommand{\shreya}[1]{\textcolor{teal}{[Shreya: #1]}}
\newcommand{\bhavya}[1]{\textcolor{magenta}{[Bhavya: #1]}}
\newcommand{\stephen}[1]{\textcolor{olive}{[Stephen: #1]}}
\newcommand{\jmh}[1]{\textcolor{olive}{[Joe: #1]}}
\newcommand{\ewu}[1]{\textcolor{red}{[wu: #1]}}
\newcommand{\dwurl}[1]{\ttt{\textcolor{blue}{\href{https://docetl.org/playground}{docetl.org/playground}}}}

\newcommand{\techreport}[1]{#1}
\newcommand{\papertext}[1]{}

\newcommand{\topic}[1]{\vspace{-3.5pt}\smallskip \smallskip \noindent{\bf #1.}}
\newcommand{\docetl}{\textsc{DocETL}\xspace}
\newcommand{\docwrangler}{\textsc{DocWrangler}\xspace}
\newcommand{\ttt}[1]{{\small \texttt{#1}}\xspace}

\definecolor{D1Color}{RGB}{255, 218, 185} %
\definecolor{D2Color}{RGB}{193, 225, 193} %
\definecolor{D3Color}{RGB}{173, 216, 230} %
\definecolor{D4Color}{RGB}{255, 239, 186} %
\definecolor{D5Color}{RGB}{221, 160, 221} %
\definecolor{D6Color}{RGB}{75, 0, 130}
\definecolor{D7Color}{RGB}{204, 204, 255} %
\definecolor{D8Color}{RGB}{152, 251, 152} %
\setlength{\fboxsep}{0.5pt}
\newcommand{\task}[2]{\colorbox{#1}{#2}} 
\newcommand{\dg}[1]{\colorbox{D8Color}{#1}} 

\definecolor{dwblue}{HTML}{007BFF}
\definecolor{dworange}{HTML}{FF9500}

\newcommand{\workflow}[1]{\textcolor{D6Color}{\ul{#1}}}

\title{Steering Semantic Data Processing With \docwrangler}

\author{Shreya Shankar$^{1\dagger}$, Bhavya Chopra$^{1\dagger}$, Mawil Hasan$^1$, Stephen Lee$^1$,\\Björn Hartmann$^1$, Joseph M. Hellerstein$^1$, Aditya G. Parameswaran$^1$, Eugene Wu$^2$}
\affiliation{%
$^1$UC Berkeley EECS, $^2$Columbia University \\
\{\url{shreyashankar, bhavyachopra, mawil0721, stephen\_lee129, hellerstein, adityagp}\} \url{@ berkeley.edu},\\\url{bjoern @ eecs.berkeley.edu}, \url{ewu @ cs.columbia.edu}\country{}}

\newif\ifanonymous
\anonymousfalse %
\ifanonymous
\else
  \thanks{$^\dagger$Co-first authors. Corresponding author: Shreya Shankar.}
\fi

\begin{abstract}
Unstructured text has long been difficult to automatically analyze at scale. Large language models (LLMs) now offer a way forward by enabling \textit{semantic data processing}, where familiar data processing operators (e.g., \ttt{map}, \ttt{reduce}, \ttt{filter}) are powered by LLMs instead of code. However, building effective semantic data processing pipelines presents a departure from traditional data pipelines: users need to understand their data to write effective pipelines, yet they need to construct pipelines to extract the data necessary for that understanding---all while navigating LLM idiosyncrasies and inconsistencies. We present \docwrangler, a mixed-initiative integrated development environment (IDE) for semantic data processing with three novel features to address the gaps between the user, their data, and their pipeline: {\em (i) In-Situ User Notes} that allows users to inspect, annotate, and track observations across documents and LLM outputs, {\em (ii) LLM-Assisted Prompt Refinement} that transforms user notes into improved operations, and {\em (iii) LLM-Assisted Operation Decomposition} that identifies when operations or documents are too complex for the LLM to correctly process and suggests decompositions. Our evaluation combines a think-aloud study with 10 participants and a public-facing deployment\techreport{ (available at \href{https://docetl.org/playground}{docetl.org/playground})} with 1,500+ recorded sessions, revealing how users develop systematic strategies for their semantic data processing tasks; e.g., transforming open-ended operations into classifiers for easier validation and intentionally using vague prompts to learn more about their data or LLM capabilities.
\end{abstract}

\ccsdesc[500]{Information systems~Data management systems}
\ccsdesc[500]{Human-centered computing~Interactive systems and tools}
\ccsdesc[500]{Computing methodologies~Artificial intelligence}

\keywords{Data Processing, Large Language Models, Human-AI Interaction}

\begin{teaserfigure}
\centering
  \includegraphics[width=0.9\linewidth]{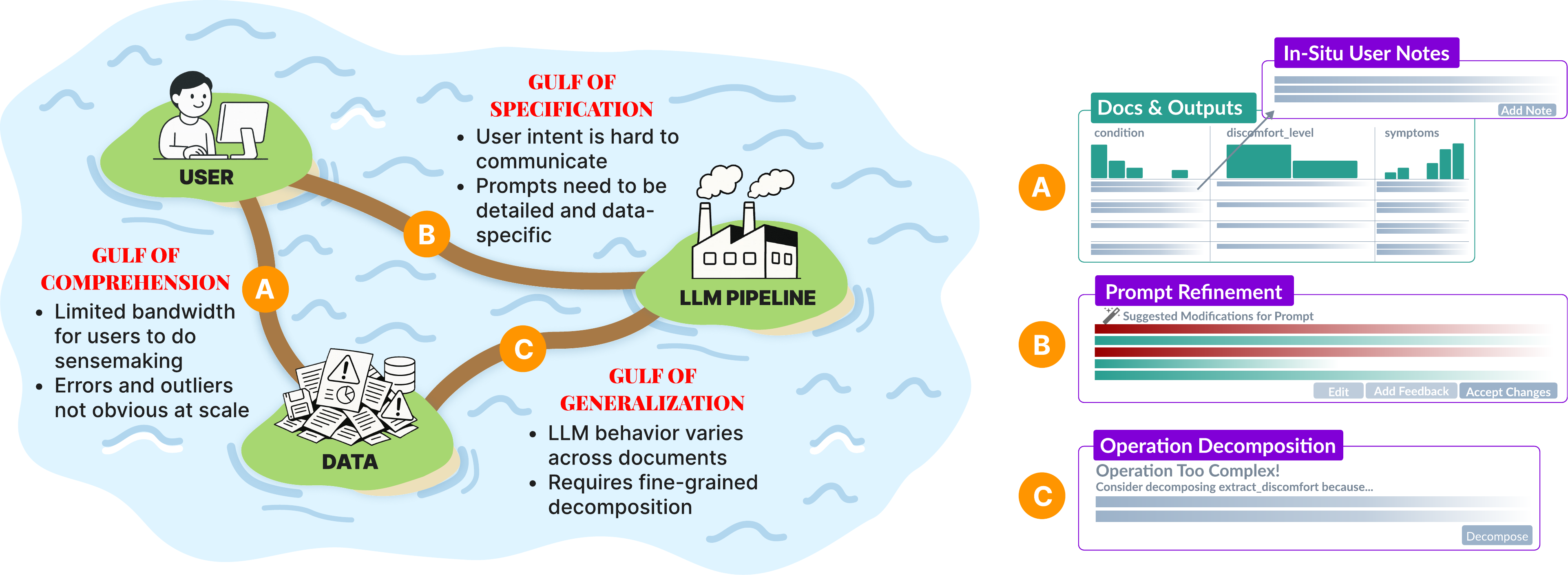}
\Description{Illustration styled as a map with islands representing the User, Data, and LLM Pipeline, separated by bodies of water symbolizing key challenges in semantic data processing. The three "gulfs"—Gulf of Comprehension (between User and Data), Gulf of Specification (between User and LLM Pipeline), and Gulf of Generalization (between Data and LLM Pipeline)—are depicted as water barriers. Arrows trace a path across the islands, labeled A, B, and C. To bridge these gulfs, DocWrangler introduces three features: in-situ user notes (A), prompt refinement (B), and operation decomposition (C), shown through corresponding UI mockups on the right.}
  \caption{Effective semantic data processing requires interaction between users, their pipeline of LLM calls, and their data (both raw documents and LLM outputs). \docwrangler contributes three novel features to address: \textcolor{dworange}{(A)} the gulf of comprehension through {\em in-situ user notes}, \textcolor{dworange}{(B)} the gulf of specification through {\em LLM-assisted prompt refinement}, and \textcolor{dworange}{(C)} the gulf of generalization through {\em LLM-assisted operation decomposition}.}
  \label{fig:cycle}
\end{teaserfigure}

\maketitle

\section{Introduction}
\label{sec:intro}

Unstructured documents---such as PDFs, reports, transcripts, and emails---are notoriously difficult to analyze automatically \cite{doan2009information}. Traditional data analysis systems like relational databases \cite{hellerstein2007architecture} and map-reduce \cite{dean2008mapreduce} typically operate on structured data (i.e., tables) and fail to effectively handle unstructured content. But large language models (LLMs) now present a way forward through {\em semantic data processing}~\cite{shankar2024docetlagenticqueryrewriting, liu2024declarative, anderson_design_2024, patel2024lotus, wang2025aop}: a paradigm where users can instruct LLMs to manipulate data through familiar data processing operators like \ttt{map}, \ttt{reduce}, \ttt{filter}, and \ttt{groupby}. For example, consider analyzing medication side effects in unstructured doctor-patient conversation transcripts. For this task, a semantic \ttt{map} could extract mentions of medications and reported side effects, followed by a semantic \ttt{reduce} to summarize effects per medication. The impact of semantic data processing could be transformative: just as traditional data analysis systems enabled structured data processing at scale and are now ubiquitous across sectors, semantic data processing has the potential to revolutionize how we work with data, but for the even-bigger realm of unstructured documents~\cite{madden2024databases, fernandez2023large}.

However, current semantic data processing systems remain far from realizing their potential. As \Cref{fig:cycle} illustrates, effective semantic data processing involves three components---the user, their pipeline of LLM operations, and the data (both raw documents and LLM outputs)---with significant challenges emerging from their interactions. Drawing on Norman's gulfs of execution and evaluation~\cite{norman1983mental}, we identify ``gulfs'' that pertain to each pairwise interaction. First, consider the {\em gulf of comprehension} between users and their unstructured data. Documents contain too much information for humans to fully process~\cite{kandel2011wrangler, lam2024concept}, and LLMs, when asked to process many documents, inevitably and unpredictably introduce mistakes or misinterpretations that users find difficult to keep track of. Second, there is a {\em gulf of specification} between users and their semantic data processing pipelines: users must first discover their true intent---often only possible after exploring sufficient data to understand what questions the data can reasonably answer---and then express this intent as pipelines of operations with corresponding LLM prompts~\cite{subramonyam2024bridging, tamkin2022task}. In both cases, users struggle with distilling their many observations about data patterns and LLM behaviors into effective specifications. Third, the {\em gulf of generalization} between the semantic data processing pipeline and data persists even when a pipeline perfectly captures user intent. Even with clear, unambiguous prompts, LLMs may fail to generalize correctly to the user's actual data, struggling with long documents~\cite{liu2024lost} or complex operations requiring simultaneous reasoning about multiple elements~\cite{kalai2024calibrated}. Complex tasks need to be decomposed into multiple LLM calls instead of one~\cite{shankar2024docetlagenticqueryrewriting, soden2020embracing, feng2024cocoa}.

\begin{figure*}
    \centering
    \includegraphics[width=0.8\linewidth]{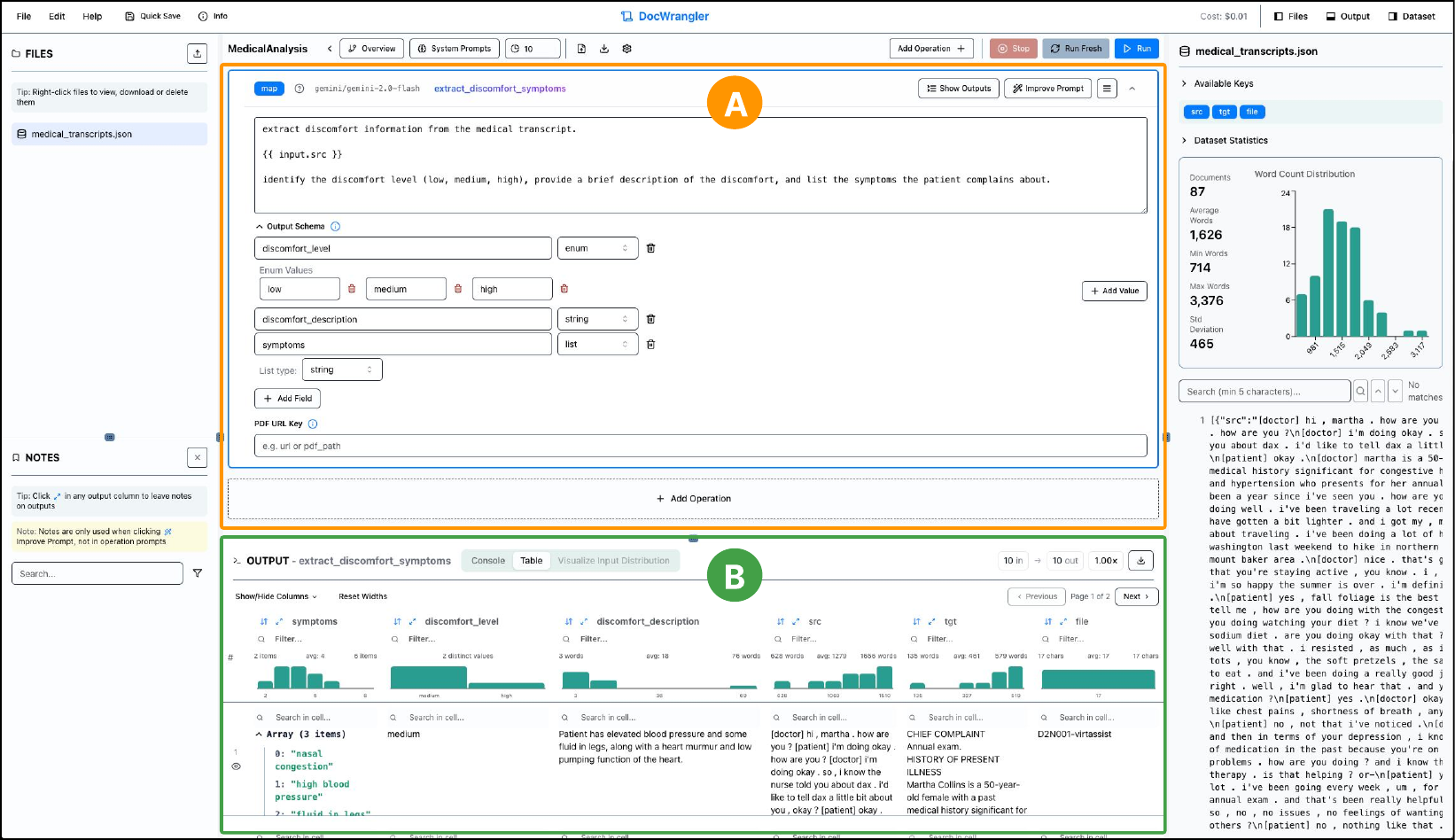}
    \Description{The figure shows a screenshot of the \docwrangler IDE interface. The layout resembles a computational notebook with multiple integrated components. The top section (labeled A) displays a notebook-style pipeline constructor with code or prompt cells and text inputs. The bottom section (labeled B) shows a spreadsheet-like data inspector displaying column headers, data visualizations in the form of small bar charts for different attributes, and sample data entries in a tabular format. The left sidebar contains a section for user notes on documents, while the right sidebar displays the raw document collection with statistics and visualizations about the document corpus. The interface follows a modern web-based IDE design with typical elements like toolbars, buttons, and dropdown menus at the top. The caption explains that this screenshot demonstrates \docwrangler's notebook-style pipeline constructor and spreadsheet-like data inspector, with user notes on documents and LLM-generated data provided in-situ, and the raw document collection displayed in the sidebars.}
    \caption{ Screenshot of the \docwrangler IDE showing our notebook-style pipeline constructor (A) and spreadsheet-like data inspector (B). User notes on documents and LLM-generated data, provided in-situ, and the raw document collection are displayed in the left and right sidebars, respectively.}
    \label{fig:maininterface}
\end{figure*}

To address the aforementioned gulfs, we present \docwrangler, a mixed-initiative integrated development environment (IDE) designed for semantic data processing (\Cref{fig:maininterface}). \docwrangler builds on established interaction patterns for transforming data: a notebook-like pipeline constructor with operation cells that can be reordered and toggled~\cite{kluyver2016jupyter}, and a spreadsheet-like output viewer with automatically-generated visualizations based on data types~\cite{kandel2011wrangler}. We additionally contribute three novel features that each address a gulf in \Cref{fig:cycle}. Our {\bf \em in-situ user notes} feature tackles the comprehension gulf by enabling users to annotate observations directly on both documents and outputs, creating a persistent, searchable record that helps track patterns across complex datasets with minimal context switching. Our {\bf \em LLM-assisted prompt refinement} feature addresses the specification gulf through an interactive interface where an LLM analyzes the pipeline, documents, outputs, and user notes to suggest more effective prompts---that the user can then tweak if they would like. Finally, our {\bf \em LLM-assisted operation decomposition} feature targets the generalization gulf by identifying when the pipeline is inadequate for the documents, using an LLM-as-judge that runs in the background~\cite{shankar_who_2024, zheng_judging_2023}. Users receive notifications highlighting detected issues along with actionable suggestions for breaking operations into more manageable steps.

To understand how \docwrangler supports semantic data processing in practice, we evaluated its use as a design probe, focusing on user needs, strategies, and workflows. Our evaluation used two complementary approaches: {\em (i)} a think-aloud study with 10 participants across different domains that provided rich qualitative insights into users' mental models and decision-making processes, and {\em (ii)} a public online deployment\techreport{ (available at \href{https://docetl.org/playground}{docetl.org/playground})} that attracted over 1,500 uses, allowing us to quantitatively analyze pipeline structures and how they changed. We observed that users develop systematic strategies for their analysis tasks, for example, by transforming open-ended operations into classifiers for easier validation. Users also employed intentionally vague prompts in \ttt{map} operations to learn more about their data, reminiscent of epistemic actions, i.e., actions taken not to directly achieve a goal but to gather information that reveals new possibilities \cite{kirsh1994distinguishing, tricaud2023revisiting}.

In summary, we contribute the following:

\begin{itemize}[nosep, left=0pt]
    \item The design and implementation of \docwrangler, an IDE for semantic data processing;
    \item Findings from a 10-person think-aloud study that demonstrate how users opportunistically align with LLMs, with rich insights into both progress and challenges; and
    \item Insights from a real-world deployment showing how semantic data processing is used across domains and areas where users need additional support.
\end{itemize}

 We first review related work in \Cref{sec:related}. We next give a more detailed background on operators in semantic data processing systems and motivate \docwrangler's design in \Cref{sec:design}. Subsequently, we describe our interface and implementation in \Cref{sec:interface}. Then, we present our study methodology in \Cref{sec:method}. We detail our findings from our user study in \Cref{sec:findings} and deployment in \Cref{sec:deployment}, and discuss implications for future system design and human-AI collaboration in \Cref{sec:discussion}.

\section{Related Work}
\label{sec:related}

In this section, we review relevant work that informs \docwrangler's development: semantic data processing systems that enable LLM-powered operations on text, then, interfaces for structured data analysis, and finally, solving general tasks with LLM pipelines.

\topic{Semantic Data Processing Systems} The goal of semantic data processing is to leverage LLMs to process large volumes of unstructured text through natural language instructions. For instance, a pipeline might use an LLM-powered \ttt{map} operation to classify sentiment in customer reviews, followed by an LLM-powered \ttt{reduce} operation that summarizes common themes for each class (i.e., positive and negative). While several systems now support semantic data processing specified through code~\cite{liu2024declarative, shankar2024docetlagenticqueryrewriting, anderson_design_2024, patel2024lotus}, users struggle to build effective pipelines with these systems~\cite{anderson_design_2024, liu2025palimpchat}. Some systems try to automatically optimize pipelines for cost~\cite{liu2024declarative} or accuracy ~\cite{shankar2024docetlagenticqueryrewriting}. However, the end-user challenges of determining what data processing intent to express and how best to express it remain unsolved, given that user preferences may depend on LLM outputs themselves~\cite{shankar_who_2024}. Moreover, automatic accuracy optimization techniques often require extensive time and computational resources to explore different decompositions and prompts~\cite{khattab_dspy_2023, shankar2024docetlagenticqueryrewriting, snell2024scaling}, making them impractical for interactive settings. 
Generally, semantic data processing systems focus on optimizing an ``inner'' computational loop that assumes well-specified tasks and clear evaluation metrics. However, real-world semantic data processing rarely begins with perfectly formed specifications, and current systems provide minimal support for the crucial human-centered ``outer'' loop of discovery and refinement. \docwrangler provides such an environment for users to author and test their semantic data processing pipelines through iterative exploration.

\topic{Interfaces for Structured Data Analysis} The history of data analysis tools provides valuable insights for building semantic data processing interfaces. We briefly cover interaction paradigms for data analysis, all of which operate instead on structured data, as semantic data processing of unstructured text is a new field.

First, {\em coding libraries}, from early ones like SAS~\cite{li2013handbook} to modern popular ones like Pandas~\cite{mckinney2011pandas}, provide data transformation capabilities through API calls. While such libraries provide high flexibility, they require users to constantly switch between programming environments and data inspection tools.\techreport{ {\em Graphical dataflow} interfaces~\cite{johnson2001labview} improved upon this workflow by visualizing processing steps through ``boxes-and-arrows'' metaphors, though they still obscure the actual data being transformed.} {\em Spreadsheets}~\cite{visicalc1979} and {\em visual analytics tools}~\cite{stolte2002polaris} instead opt for direct manipulation for editing and visualizing data respectively, making data analysis more accessible to non-programmers. \techreport{Over the decades, research has consistently shown that ``always-on'' visualizations can complement data processing tools, helping users discover insights and validate transformations faster~\cite{lee2021lux, kandel2011research, shrestha2021unravel}, especially when users can directly interact with visualizations~\cite{siddiqui2016effortless}.} 

Hybrid approaches aim to combine the best aspects of the previous paradigms.\techreport{ Systems like Potter's Wheel~\cite{raman2001potter} combine interactive interfaces with underlying domain-specific languages for data transformation.} {\em Programming-by-example} approaches~\cite{gulwani2016programming, yessenov2013colorful} infer transformations from user demonstrations in limited settings (e.g., regex-based transformations)\techreport{, though for enterprise data processing they require many examples to get started, and are difficult to validate and generalize.} {\em Predictive interaction}~\cite{heer2015predictive} introduced a mixed-initiative approach to data transformation through a ``guide/decide'' loop, where systems present suggested transformations based on user selections, while also letting users inspect data to see which transforms are necessary~\cite{kandel2011wrangler}.

More recently, LLMs have transformed how users interact with data, though we are still in the early stages of this evolution.
The first generation of LLM-based tools primarily created natural language interfaces for {\em structured data} tasks: specifically, conversational assistants that generate SQL or Pandas code~\cite{gu2024data, hassani2023role, chopra2023conversational, zhao2024chat2data, setlur2022you}. 
Steering these AI-generated pipelines is difficult~\cite{mcnutt2023design}: LLMs often misinterpret user intent or produce incorrect code, requiring specialized interfaces for correction. 
\citet{xie2024waitgpt} propose visualizing pipelines as directed acyclic graphs, reminiscent of dataflow GUIs, while \citet{kazemitabaar2024improving} propose a notebook-style interface that breaks complex workflows into discrete, inspectable steps. These environments are code-centric: they prioritize code review over data inspection, overlooking the well-established value of integrated data visualization~\cite{lee2021lux, siddiqui2016effortless, wu2020b2}. Since data is messy and transformations rarely work on the first try, users need to switch between writing code and reviewing outputs to catch anomalies and spot data that still needs processing.

Moreover, perhaps because they focus on structured data, prior work treats LLMs merely as code generators rather than semantic operators in their own right. In semantic data processing, LLMs aren't just writing scripts in a traditional data processing language, they provide entirely new black-box capabilities for unstructured data transformations. This creates a new challenge: while traditional data processing code can often be validated through static analysis or test cases, LLM behavior in semantic operations is inherently uncertain and context-dependent. Effective interfaces for semantic data processing must therefore enable refinement of both the pipeline structure {\em and} individual operations while keeping data continuously visible throughout the process. 

\topic{LLM Workflow Development and Validation} As LLMs tackle increasingly complex problems, like semantic data processing, users need multi-step workflows to express their goals. Effective interfaces for general problem-solving with LLMs should support the entire process---from task decomposition to output refinement---building on established research in mixed-initiative interfaces~\cite{horvitz1999principles}, collaborative AI~\cite{amershi2019guidelines}, and interactive machine learning~\cite{dudley_review_2018}.

Task decomposition has deep roots in both human and AI systems. Crowdsourcing research introduced strategies for breaking down complex tasks, some of which has been successfully applied to LLMs~\cite{grunde-mclaughlin_designing_2024, parameswaran2023revisiting}, and cognitive science has studied human approaches to problem decomposition~\cite{newell1972human}. Recent interfaces apply these insights to convert natural language into executable sub-tasks~\cite{kazemitabaar2024improving, zhang_chainbuddy_2024}, but users still struggle with information overload when examining LLM-generated workflows~\cite{zamfirescu2025beyond, kazemitabaar2024improving, chen2024need}. This points to a critical need for interfaces that help users make sense of these complex workflows. 
Some tools address pipeline sensemaking by enabling interactive prompt experimentation and output inspection~\cite{arawjo2024chainforge, wu2022ai}, although these are not designed for data processing. Unlike general LLM pipelines, semantic data processing faces steeper challenges in bridging the gulfs shown in \Cref{fig:cycle}. The heterogeneous nature and scale of unstructured documents make both comprehension and LLM generalization more difficult. Separately, while semantic data processing is underexplored, we are starting to see ``point'' solutions address specific semantic data processing applications; e.g., LLooM's interface for concept induction tasks~\cite{lam2024concept}. However, we lack {\em general-purpose} interfaces for semantic data processing across diverse document and operator types. Designing such interfaces is not straightforward, as users encounter the ``gulf of envisioning''~\cite{subramonyam2024bridging}---the cognitive gap between having a goal and translating it into effective LLM instructions---while also understanding how to evaluate whether the output meets their original intentions.\footnote{Note that for semantic data processing,~\citet{subramonyam2024bridging}'s ``gulf of envisioning'' encompasses all the gulfs presented in our model in \Cref{fig:cycle}.}

Even when a complex task is expressed as a pipeline of well-scoped operations, validating and debugging LLM outputs remains challenging~\cite{liu2022design, dang2022prompt, arawjo2024chainforge}. Users face high cognitive load from constantly switching between prompting and evaluation~\cite{tankelevitch2024metacognitive, strobelt2022interactive}, and developers struggle with LLMs' unpredictability---fixing one issue often creates others~\cite{zamfirescu2023herding}. Some tools address these challenges through automated evaluation, often employing an {\em LLM-as-judge} methodology where an LLM evaluates outputs against specific criteria~\cite{zheng_judging_2023, szymanski2025limitations, shankar_spade_2024, shankar_who_2024, kim2024evallm, ma2024you}.  \techreport{These automated evaluation approaches, which focus on validating individual outputs, have not been considered in the context of semantic data processing, where users need to assess both individual results and aggregate patterns across entire datasets. Moreover, in semantic data processing, different types of operations require distinct evaluation approaches---for example, assessing the accuracy of information extraction is different from assessing the quality of a summary. Specialized visualizations~\cite{gero2024supporting, suh2023sensecape} can help users validate and understand LLM behavior, though again, these have not been applied to semantic data processing.}  Additionally, identifying errors in LLM outputs is only half the battle. Users also need tools to refine pipelines based on observed patterns. We build on prior work that proposes automatic prompt refinement for image generation~\cite{brade2023promptify, almeda2024prompting}: for semantic data processing, prompts may need to change in ways beyond specific attributes (e.g., subject or style) as users discover their true question, potentially shifting the entire task direction.

\topic{Summary} Overall, prior work points to a clear challenge: semantic data processing systems lack effective interfaces for users to iteratively develop, validate, and refine their pipelines. Prior work also sheds light into what we should care about when building an interface for semantic data processing workflows: treating them as data transformation workflows first and foremost, while simultaneously providing tools for effective prompt engineering, enabling users to validate both individual outputs and aggregate patterns across documents, and helping users navigate the gulfs  proposed in \Cref{fig:cycle}. \docwrangler embodies the aforementioned principles as an IDE, while preserving users' agency throughout~\cite{horvitz1999principles, amershi_guidelines_2019}.

\section{\docwrangler Design}
\label{sec:design}

We now describe the foundation and design principles of \docwrangler. While our interface builds on \docetl~\cite{shankar2024docetlagenticqueryrewriting}, an open-source semantic data processing system, as the backend, we could have equally well chosen to build \docwrangler on top of other semantic data processing systems ~\cite{anderson_design_2024, patel2024lotus, liu2024declarative, wang2025aop}. We first provide background on \docetl, then present our design goals based on an analysis of user needs and challenges in semantic data processing.

\subsection{\docetl Background and Example}

\docetl is a declarative framework for building semantic data processing pipelines, where many of the unit data processing operations are executed by LLMs. 
Each LLM-powered operator is defined through two components: a natural language {\em prompt} that specifies what the operation should do, and an {\em output schema} that determines the structure of data the LLM should generate. Inputs to \docetl are documents, represented as JSON collections of key-value pairs, allowing flexible handling of both structured and unstructured content. An entirely unstructured document (like an email or news article) can be represented as a single key-value pair (e.g., \(\{\text{"content"}: \text{"full text here..."}\}\)). \docetl offers multiple semantic operators, including: \ttt{map} (applies prompts to individual documents), \ttt{filter} (removes documents that don't meet specified criteria), \ttt{reduce} (processes document groups collectively), and \ttt{resolve} (performs entity resolution and canonicalization across documents). These operations leverage Jinja templates in their prompts~\cite{nipkow2003jinja}, enabling users to reference specific document key-value pairs. As the pipeline executes, each operation enriches documents with new key-value pairs according to its output schema, allowing subsequent operations to build upon earlier results.

To illustrate how these components work together in practice, consider a semantic processing pipeline analyzing student course reviews to identify common themes of complaints, with supporting evidence. The pipeline might consist of three operations:

\begin{enumerate}[nosep, left=0pt]
    \item  A \ttt{map} operation that processes each review (represented as a separate JSON document) individually, with a prompt asking the LLM to {\em ``Extract complaint themes and their supporting quotes from this review''} and an output schema defining two new attributes, \ttt{themes} (an array of strings) and \ttt{supporting\_quotes} (an array of strings corresponding to each theme).
    \item A \ttt{resolve} operation that semantically de-duplicates themes across all reviews (e.g., recognizing that {\em``professor talks too fast''} and {\em ``professor speaks quickly''} represent the same underlying complaint), with a comparison prompt to {\em ``Consider if these two themes are similar and return \ttt{True} if so''}; and a resolution prompt to {\em ``Output a single theme that best represents these themes.''}
    \item A \ttt{reduce} operation that groups reviews by theme and generates a summary report for each theme, with a prompt like {\em ``Summarize the common sentiments and representative quotes for this theme,''} and an output schema for the new \ttt{summary} attribute.
\end{enumerate}

Although a system like \docetl can relieve users of the low-level execution details (e.g., orchestrating all LLM calls associated with such a pipeline), constructing effective pipelines remains challenging because users often cannot fully specify their analytical intents in advance. For example, what the LLM ``believes'' is a complaint may not be the type or granularity of complaint the user is interested in, or the themes extracted might combine issues the user would prefer to see separately. 

\subsection{\docwrangler Design Goals}

To better understand the challenges users face when building LLM-powered data processing pipelines, we analyzed messages and feedback from the \docetl Github and Discord community\footnote{\url{https://discord.gg/fHp7B2X3xx}} (400+ members) and drew insights from prior work (described in \Cref{sec:related}). 

We observed that users' workflows in semantic data processing typically follow three phases that we call the {\bf ``Three I's.''} Users cycle between {\bf initializing} their pipelines by defining operations, then {\bf inspecting} outputs to understand results, and {\bf improving} their pipelines based on these insights. We organize user pain points across these phases. First, there are {\bf initialization} challenges. Writing effective operations requires predicting how well an LLM will interpret a user's data processing intent. This challenge creates a premature commitment problem \cite{blackwell2003notational}---users invest time crafting operations without knowing if they'll work, often wasting effort when outputs fail to meet expectations. Users expressed a desire for assistance in writing the pipeline, as well as ``preview'' functionality to validate their approach before committing to full execution. 

Next, there are {\bf inspection} challenges. After initializing, users struggle to validate whether LLM behavior across documents aligns with their expectations for each operation. The volume of data makes review of all documents and LLM outputs impractical. Users often don't know when they're ``losing the forest for the trees'' or vice versa when reviewing outputs. Users expressed wanting to review intermediate operation outputs, and some mentioned that they exported outputs as spreadsheets to look at results as different views (row by row, all at once in a spreadsheet, or side by side) in an ad-hoc manner. Fragmenting the validation workflow across multiple tools creates additional cognitive load for users~\cite{czerwinski2004diary, kandel2012enterprise}.

Third, there are {\bf improvement} challenges. Users struggle to externalize patterns---both positive and negative---observed across multiple documents. Some users took notes in spreadsheets or a separate text file as a way to organize their insights, but they struggled to translate their notes into concrete pipeline modifications. Prior work also observes that users struggle to prompt LLMs to edit pipelines~\cite{kazemitabaar2024improving}. Moreover, users did not know how to improve pipelines when the task was ``too hard'' for the LLM, requiring operation decomposition, or a redistribution of tasks between LLM-powered and code-powered operators.

We also turned to prior work in structured data processing interfaces to understand what users might need in an semantic data processing interface. Visualization research demonstrates that users need both overview and detail views to efficiently make sense of complex datasets~\cite{shneiderman2003eyes}\techreport{---i.e., showing aggregates first helps users identify patterns, while enabling drill-down into specific examples supports verification~\cite{stolte2002polaris, lee2018case}}. Data wrangling tools demonstrate improved user productivity when users can see and edit data transformation code directly~\cite{kandel2011wrangler}. Mixed-initiative interfaces highlight the importance of preserving user agency, ensuring system observability, and reducing context-switching costs when supporting complex, iterative tasks~\cite{horvitz1999principles, amershi_guidelines_2019, shneiderman2020human}. 

Overall, we formulated five design goals for an effective semantic data processing interface:

\begin{itemize}[nosep, left=0pt]
\item {\bf \dg{D1}. Scaffold Pipeline Initialization}: Help users create and configure operations with minimal friction, with built-in guidance and quick experimentation.
\item {\bf \dg{D2}. Facilitate Efficient Data Inspection and Notetaking}: Enable users to validate inputs and outputs individually and in aggregate, while supporting note-taking to capture insights and patterns.
\item {\bf \dg{D3}. Guide Pipeline Improvement}: Offer assistance for translating user feedback into effective pipeline modifications, both at the individual operation level (e.g., prompt improvements) and pipeline level (e.g., operation decomposition).
\item {\bf \dg{D4}. Maintain End-to-End Observability}: Ensure transparency into transformation logic at each pipeline step (e.g., inputs, outputs, LLM prompts).
\item {\bf \dg{D5}. Minimize Context Switching}: Integrate all essential analytical capabilities within a unified interface, minimizing the need for external tools (e.g., spreadsheets, custom scripts, AI assistants like ChatGPT).
\end{itemize}

\section{\docwrangler System}
\label{sec:interface}

\begin{table}[t]
\footnotesize
\begin{tabular}{p{1cm}p{5.5cm}l}
\toprule
\textbf{Phase} & \textbf{Feature} & \textbf{Goals} \\
\midrule

\multirow{4}{*}{Initialize} 
  & Pipeline editor with operation cards that can be toggled and reordered & \dg{D1}, \dg{D4} \\
  & Custom cards for each operator type with live syntax validation & \dg{D1} \\
  & Visual document flow between operations to show transformation paths & \dg{D1}, \dg{D4} \\
  & Execution on sampled documents for quick iteration & \dg{D1} \\

\midrule

\multirow{6}{*}{Inspect} 
  & Spreadsheet-style viewer with automatic visualizations per column & \dg{D2} \\
  & Support for column resizing, filtering, sorting, and search & \dg{D2}, \dg{D5} \\
  & Per-document prompt viewer to inspect exact LLM inputs & \dg{D4} \\
  & Detailed document viewer with keyboard navigation and side-by-side comparison & \dg{D2}, \dg{D5} \\
  & Output inspection available at any pipeline stage, including intermediates & \dg{D2}, \dg{D4} \\

\midrule

\multirow{6}{*}{Improve} 
  & In-situ note-taking system for annotating and categorizing output issues & \dg{D3}, \dg{D5} \\
  & Persistent notes sidebar with filtering and search across iterations & \dg{D3}, \dg{D4} \\
  & AI-assisted Prompt Refinement interface driven by user notes & \dg{D3}, \dg{D5} \\
  & Automatic operation decomposition with explanations and plan diffs & \dg{D3}, \dg{D4} \\
  & Embedded AI assistant for help with prompts, templates, and workflow guidance & \dg{D1}, \dg{D3}, \dg{D5} \\

\bottomrule
\end{tabular}
\Description{The table has three columns: Phase, Feature, and Goals. The table is divided into three main phases: Initialize, Inspect, and Improve. Under Initialize phase: features include pipeline editor with toggleable operation cards (goals D1, D4), custom cards for each operator type with live syntax validation (D1), visual document flow between operations (D1, D4), and execution on sampled documents (D1). Under Inspect phase: features include spreadsheet-style viewer with visualizations (D2), column resizing/filtering/sorting (D2, D5), per-document prompt viewer (D4), detailed document viewer with keyboard navigation (D2, D5), and output inspection at any pipeline stage (D2, D4). Under Improve phase: features include in-situ note-taking system (D3, D5), persistent notes sidebar (D3, D4), AI-assisted Prompt Refinement interface (D3, D5), automatic operation decomposition (D3, D4), and embedded AI assistant for workflow guidance (D1, D3, D5). Each feature is linked to specific design goals labeled D1 through D5.}
\caption{Key features in \docwrangler, grouped by phase of semantic data processing, and linked to design goals.}
\vspace{-20pt}
\label{tab:features}
\end{table}

\begin{figure*}
    \centering
    \includegraphics[width=0.8\linewidth]{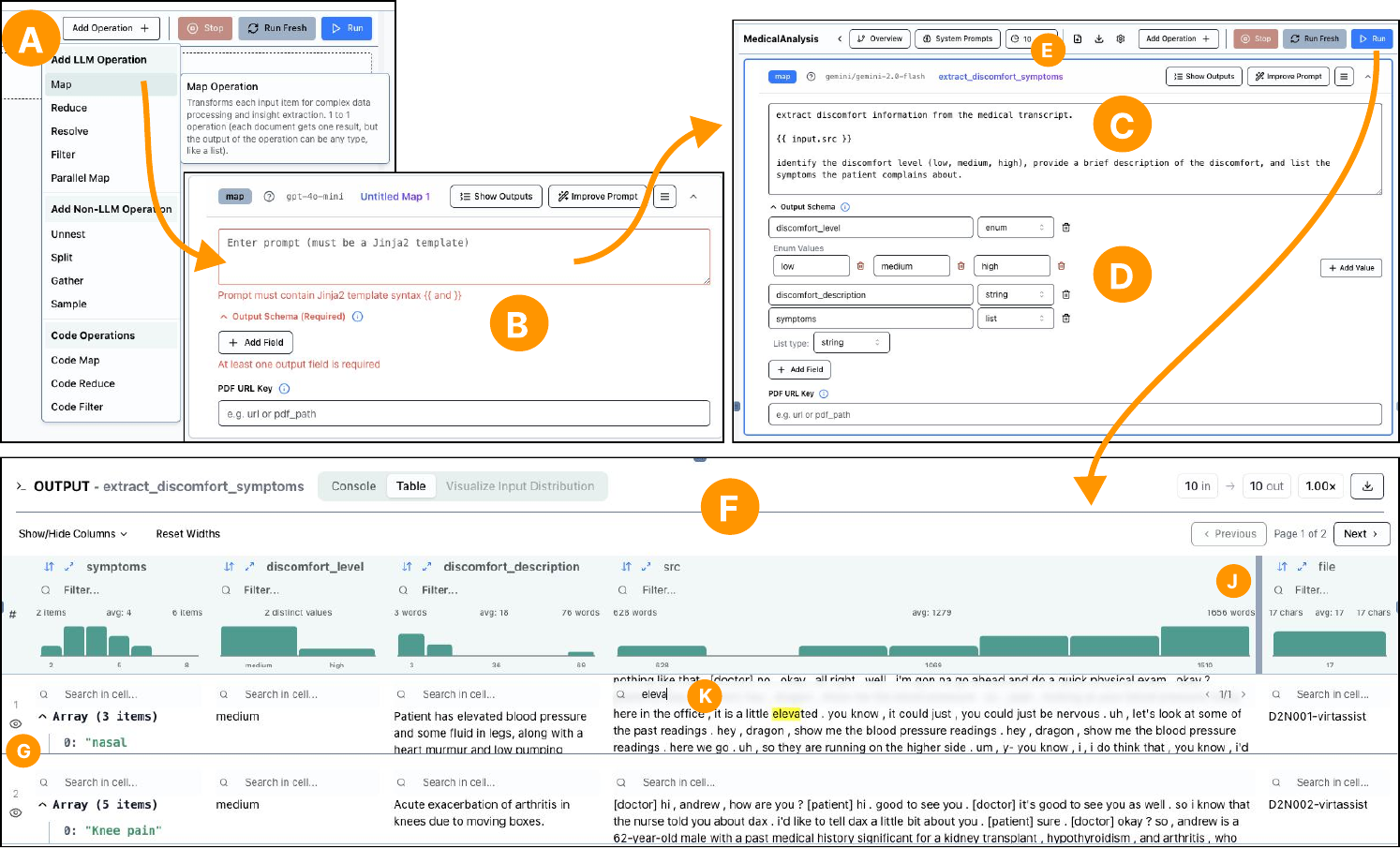}
    \Description{Figure 3 presents the DocWrangler workflow for analyzing patient discomfort from medical transcripts. The image shows different interface components labeled A through K. The diagram includes a few connecting arrows between some components to show workflow progression: from A (operation selection) to B (map operation configuration), then to C (prompt writing) and eventually to F (results table). The remaining labeled elements (D-E, G-K) highlight specific interface features rather than sequential steps.}
    \caption{Workflow for analyzing patient discomfort from medical transcripts (\dg{D1}; \dg{D2}). (A) The user adds a new operation via dropdown menu. (B) A \ttt{map} operation card appears with syntax validation. (C) The user writes a prompt for extracting discomfort information. (D) The user defines an output schema with three attributes to be extracted: discomfort level, description, and symptoms. (E) Before running the full dataset, they sample 10 documents. (F) Results appear in an interactive table. (G) Eye icons reveal exact LLM prompts for each document. (H) Column headers visualize attribute distributions. (I) Sort buttons reorder documents. (J) Adjustable column widths help focus on specific content. (K) Search functionality helps validate extractions.}
    \label{fig:initworkflow}
    \vspace{-10pt}
\end{figure*}

\begin{figure}
    \centering
    \includegraphics[width=0.85\linewidth]{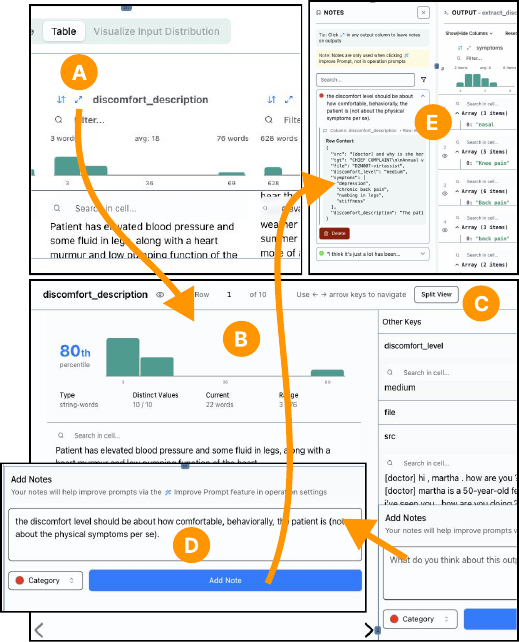}
    \Description{The figure contains five labeled interface screenshots, A to E, showing the in-situ user notes workflow in DocWrangler. A shows a user selecting an attribute called ``discomfort\_description'' from a data table. B displays a document viewer panel with a bar chart of attribute statistics and annotated text. C shows a split-screen view with documents on one side and keys like ``discomfort\_level'' on the other. D shows a note entry box where the user writes that discomfort level should refer to emotional, not physical, comfort. E shows the saved note listed in the sidebar under the ``NOTES'' section. Arrows guide the viewer from attribute selection to note creation and persistence.}
    \caption{\workflow{\bf In-situ user notes} feature in \docwrangler (\dg{D2}). (A) User selects attribute to inspect. (B) Document viewer dialog shows attribute statistics and enables in-situ notes. (C) User could inspect documents side-by-side with split-screen view, if they want. (D) User adds a note. (E) Notes persist in the main interface sidebar.}
    \vspace{-10pt}
    \label{fig:feedbackworkflow}
\end{figure}

\begin{figure*}
    \centering
    \includegraphics[width=0.85\linewidth]{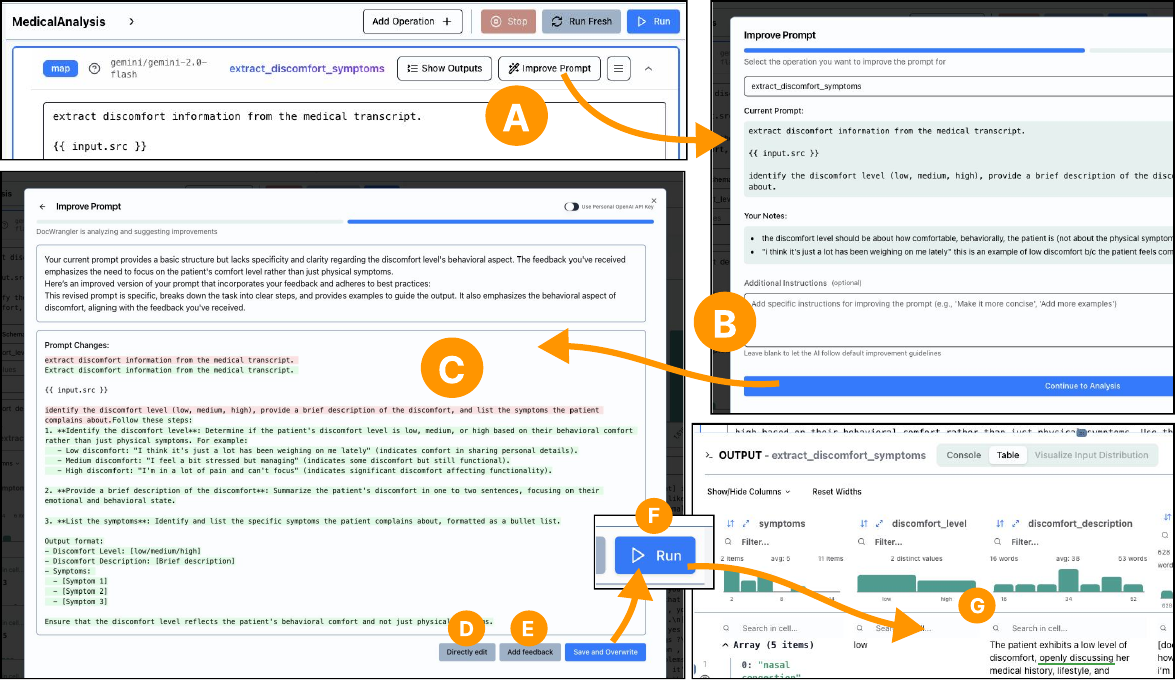}
    \Description{The figure contains seven labeled interface screenshots arranged to show a prompt refinement workflow in the ``medical analysis'' pipeline. Screenshot A shows a code editor interface with a prompt and an ``Improve Prompt'' button. B shows a dialog box with the current prompt and user feedback. C presents an AI-suggested improved prompt with detailed instructions. D and E show the same prompt box with behavioral examples added and buttons for editing or accepting. F shows the interface with the ``Run'' button highlighted. G displays the resulting bar chart with improved discomfort-level annotations. Orange arrows guide the viewer through the A--G sequence.}
    \caption{\workflow{\bf Prompt Refinement} workflow (\dg{D3}). (A) User initiates improvement via ``Improve'' button. (B) A dialog opens, showing the current operation and the user's relevant in-situ notes. (C) AI suggests improved prompt addressing notes (focusing on behavioral discomfort). (D) Improved prompt includes behavioral indicator examples. (E) User can edit the suggestion directly, or instruct the AI to edit it. (F) User runs updated operation. (G) Results show a more accurate discomfort distribution.}
    \label{fig:improvementworkflow}
\end{figure*}

\begin{figure*}
    \centering
    \includegraphics[width=0.85\linewidth]{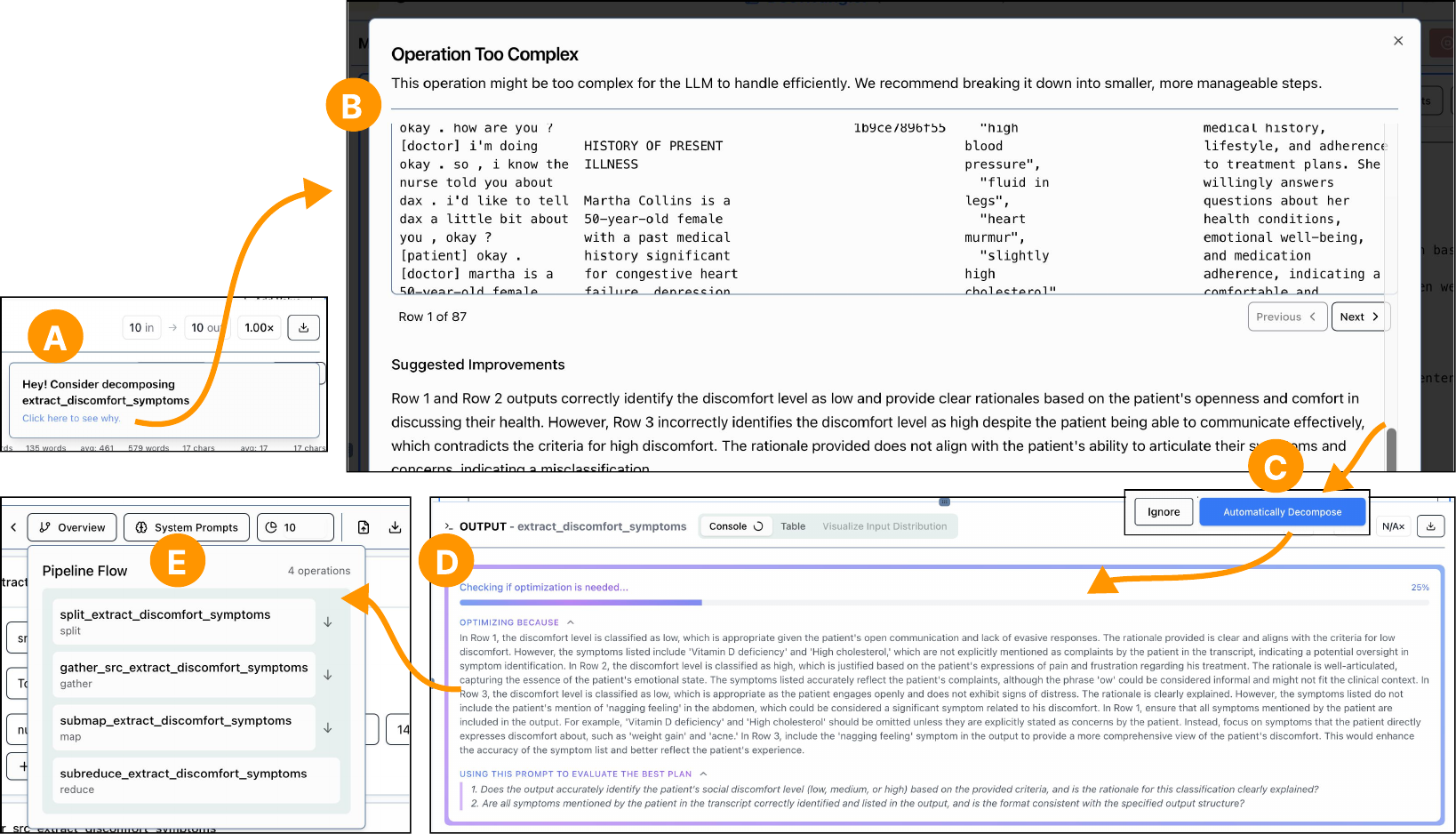}
    \Description{The figure contains five labeled screenshots, A to E, arranged to illustrate the operation decomposition workflow. Screenshot A shows a notification suggesting decomposition of a complex operation. B displays a dialog with highlighted text and rationale explaining the challenge of processing the input. C shows two buttons, ``Ignore'' and ``Automatically Decompose,'' allowing the user to proceed. D displays a detailed explanation of the decomposition logic and progress bar. E shows the updated pipeline view with four new decomposed operations listed, connected by arrows to show sequential flow.}
    \caption{The \workflow{\bf Operation Decomposition} workflow (\dg{D3}). (A) A notification suggests decomposing a complex operation, simultaneous extraction of discomfort and symptoms for long documents. (B) A dialog explains why the operation exceeds LLM capabilities, with examples of inconsistent outputs. (C) User can automate decomposition or ignore. (D) System shows real-time optimization process with reasoning (e.g., evaluation criteria for determining the best decomposition). (E) The resulting pipeline splits documents into chunks, executes the original operation on each chunk, and has another operation to unify results.}
    \label{fig:decompositionworkflow}
    \vspace{-10pt}
\end{figure*}

We present \docwrangler, an integrated development environment (IDE) for semantic data processing. First, we provide an overview of the solution and how it addresses our design goals. Then, we present an example usage scenario that walks through our features. Finally, we describe technical implementation details.

\subsection{Overview of Solution}

\docwrangler's interface (\Cref{fig:maininterface}) integrates a file and notes viewer, pipeline editor, and dataset viewer in a unified workspace. To {\bf scaffold pipeline initialization (\dg{D1})}, we developed a notebook-inspired~\cite{kluyver2016jupyter} editor with interactive operation cards featuring syntax validation, drag-and-drop reordering, and the ability to toggle operations on and off (\Cref{fig:maininterface}A). Once initialized, we allow users to run pipelines on samples to reduce response time, and we cache all intermediate results for future reuse. After execution, to {\bf inspect outputs efficiently (\dg{D2})}, our spreadsheet-like viewer, as shown in \Cref{fig:maininterface}B provides automated visualizations of document attributes as columns, along with sorting, filtering, resizing, and search capabilities. Our design is reminiscent of data wrangling interfaces that also show high-level summaries~\cite{kandel2011wrangler}; however, most LLM-generated attributes are unstructured text, so we show histograms of word or character counts as appropriate. While not depicted in \Cref{fig:maininterface}, \docwrangler also has a detailed document inspector that allows users to examine LLM outputs and source documents row by row, with side-by-side comparison, as will be described in \Cref{sec:interface-example-usage}.

Then, to {\bf guide pipeline improvement (\dg{D3})}, we introduce three novel features: {\em (i)} \workflow{\bf In-Situ User Notes} enables users to capture observations directly during inspection of documents and LLM outputs, with notes persisting across pipeline runs; {\em (ii)} \workflow{\bf Prompt Refinement} suggests improved versions of prompts, based on in-situ notes, through a conversational interface that supports both direct editing and AI-assisted revisions; and {\em (iii)} \workflow{\bf Operation Decomposition} proactively identifies when operations exceed LLM capabilities and offers automatic restructuring with explanations. Additionally, to {\bf maintain end-to-end observability (\dg{D4})}, we visualize document flow between operations, provide access to exact LLM prompts, and make intermediate outputs and suggested \workflow{\bf Operation Decomposition} feedback inspectable. Finally, to {\bf minimize context switching (\dg{D5})}, we integrated all analytical capabilities within a single interface and, following recent systems~\cite{zhang_chainbuddy_2024, kazemitabaar2024improving}, included a context-aware AI assistant to help users navigate the system. \Cref{tab:features} summarizes the key features of \docwrangler across each of the ``three I's'' and how they map to our design goals.

\subsection{Feature Walk-Through}
\label{sec:interface-example-usage}

We present our features by illustrating an example \docwrangler scenario, where, given a corpus of doctor-patient conversation transcripts, a medical data analyst wants to analyze how emotionally comfortable and open patients are during their visits, and how this varies by symptom reported.

\subsubsection{Initialization Phase (\dg{D1})}The analyst begins by uploading the collection of doctor-patient transcripts through the upload interface (top left of \Cref{fig:maininterface}). They can inspect their raw data in the dataset viewer on the right, which provides an overview through statistics about the transcript collection, including document count (here, 87), word count distributions, and existing metadata fields (\dg{D4}, \dg{D5}). With this initial understanding of their data, the analyst creates their first operation by clicking the ``Add Operation'' button (\Cref{fig:initworkflow}A), selecting a \ttt{map} type operation, which will create new attribute(s) for each document. While defining the operation (\Cref{fig:initworkflow}B, \Cref{fig:initworkflow}C), they write a prompt instructing the LLM to {\em ``Extract discomfort information from the medical transcript. \{\{ input.src \}\} Identify the discomfort level (low, medium, high), provide a brief description of the discomfort, and list the symptoms the patient complains about.''} The analyst then defines three new attributes for the LLM to populate (\Cref{fig:initworkflow}D). Before committing to processing their entire dataset, they enable document sampling (10 documents; \Cref{fig:initworkflow}E) and click ``Run'' to quickly test their pipeline. 

\subsubsection{Inspection Phase (\dg{D2})} The analyst reviews results in the table view (\Cref{fig:initworkflow}F), where each row represents a document and columns show attributes. Hovering over eye icons for each row (\Cref{fig:initworkflow}G) reveals the exact LLM prompt, enabling verification that the Jinja template was specified correctly (\dg{D4}). Noticing the unexpected distribution of discomfort levels in column headers (\Cref{fig:initworkflow}H)---mostly ``medium'' or ``high''---the analyst investigates by sorting documents (\Cref{fig:initworkflow}I), resizing columns (\Cref{fig:initworkflow}J), and scanning descriptions. This reveals a misalignment: LLM outputs are focused on patients' physical symptoms rather than the analyst's intended measure of patient openness and comfort level during the visit---which the analyst confirms by searching for specific terms in source documents (\Cref{fig:initworkflow}K).

\subsubsection{Improvement Phase (\dg{D3})} After reviewing outputs in the table, the analyst selects an attribute for closer examination (\Cref{fig:feedbackworkflow}A), opening a full-screen dialog with document navigation capabilities (\Cref{fig:feedbackworkflow}B). Using the \workflow{\bf In-Situ User Notes} feature, the analyst adds a note in the inspector panel (\Cref{fig:feedbackworkflow}D): {\em ``the discomfort level should be about how comfortable, behaviorally, the patient is (not about the physical symptoms per se)''} and tags it as red.\footnote{Users can color-code notes (e.g., red for critical issues, green for positive observations) similar to qualitative analysis tools~\cite{MAXQDA2022, williams2019art}.} After reviewing multiple documents, they return to the main interface where all notes are accessible, searchable, and filterable (\Cref{fig:feedbackworkflow}E).

The analyst initiates the \workflow{\bf Prompt Refinement} workflow by clicking ``Improve'' in the operation card (\Cref{fig:improvementworkflow}A), opening a dialog with their current prompt and any relevant notes (\Cref{fig:improvementworkflow}B). After providing any optional additional instructions and clicking ``Continue to Analysis,'' the user gets an improved prompt emphasizing behavioral aspects of discomfort (\Cref{fig:improvementworkflow}C), visualizing changes between versions. The analyst can directly edit the suggestion (\Cref{fig:improvementworkflow}D) or request further AI modifications (\Cref{fig:improvementworkflow}E) before saving.

\subsubsection{More Iteration} After running the pipeline again (\Cref{fig:improvementworkflow}F), the analyst notices improved discomfort level distributions (\Cref{fig:improvementworkflow}G), with many documents now correctly classified as ``low'' discomfort. The analyst moves on to analyzing the extracted symptoms. While they are inspecting symptom outputs, \docwrangler notifies the user (\Cref{fig:decompositionworkflow}A) that the operation may be too complex (\dg{D3}). Clicking on the notification triggers the the \workflow{\bf Operation Decomposition} feature. A dialog appears, showing examples of incorrect LLM results when handling both discomfort assessment and symptom extraction simultaneously (\Cref{fig:decompositionworkflow}B). The analyst clicks ``Automatically Decompose'' (\Cref{fig:decompositionworkflow}C), and the system transparently streams its accuracy optimization process (\Cref{fig:decompositionworkflow}D), evaluating different candidate plans with LLM-as-judge evaluators~\cite{zheng_judging_2023} (\dg{D4}). After a few minutes, \docwrangler then shows a visualization of the restructured pipeline (\Cref{fig:decompositionworkflow}F) that processes the same task but divides the work across multiple operations---first breaking the data into manageable chunks with a \ttt{split} operation, processing each chunk separately with the original \ttt{map} operation, and then using a \ttt{reduce} operation to unify the results. \papertext{With the decomposed pipeline now reliable, the analyst can proceed with their pipeline. See \Cref{app:extended} for an extended feature walk-through, showcasing the creation of other operations and usage of the AI assistant feature.}

\subsection{Implementation Details}
\label{sec:interface-implementation}

Here, we describe \docwrangler's implementation details. \docwrangler is built with Next.js and TypeScript on the frontend (styled with TailwindCSS), and a Python-based FastAPI backend. The frontend uses Monaco Editor for writing operations and ReCharts (a React wrapper around D3) for visualizations. The backend compiles the visual pipeline into \docetl's execution format and processes documents in parallel, streaming real-time updates via WebSockets. Pipeline definitions, user notes, and datasets are stored on disk (or S3 for the public deployment). User notes are additionally cached in browser storage for faster querying. To optimize performance, we use Python’s \ttt{diskcache} library for caching operation outputs, keyed by document ID and operation sequence hash. Only modified operations and downstream steps are recomputed. The output inspector automatically generates visualizations based on attribute type~\cite{munzner2014visualization, ware2019information}.  For numerical attributes, we render 7-bin histograms. For boolean attributes, we use 2-bin bar charts. For string data, we first assess whether it is categorical (fewer than 50\% unique values). If so, we display bar charts of the top 7 values; if not, we show word counts for multi-word outputs and character counts for single-word outputs. All charts use virtual scrolling to support large datasets.

\docwrangler is containerized via Docker for open-source use, and hosted on Modal Labs’ cloud platform for the public version. Users provide their own LLM API keys. In the following paragraph, we describe implementation details for the AI-assisted features.

\topic{Assisted Prompt Refinement} Our \workflow{\bf Prompt Refinement} feature relies on a representation of user notes. We store user notes as a list of tuples, where a note is a tuple containing: operation identifier, attribute name, free-text comment, and an optional category tag for color-coding. Our system uses a conversational AI interface but presents a specialized revision UI rather than a traditional chat (\Cref{fig:improvementworkflow}E). The interface displays only the latest AI-suggested prompt, highlighting the diff, while maintaining conversation context in the background. When users initiate prompt refinement, \docwrangler gathers in-situ user notes relevant to the operation and the corresponding documents, then sends an AI assistant, powered by gpt-4o-mini, a message containing the current prompt, output schema, a sample of documents, in-situ user notes and corresponding documents, and basic prompt engineering guidelines (i.e., be clear, unambiguous, and provide few-shot examples if possible). We also instruct the AI assistant to return structured content with new prompts enclosed within \ttt{<prompt>} and \ttt{</prompt>} tags and schema changes within \ttt{<schema>} and \ttt{</schema>} tags for easy parsing. The AI's response is streamed in real-time.

Whenever users edit a prompt directly or provide feedback to the AI assistant, we append a new message to the conversation history. For direct edits, we create a canonical message recording that the user changed X to Y; for feedback, we include the user's instruction to the LLM. All revisions are managed in a tree structure, visualized as an interactive diagram, where each node represents a prompt version. This diagram appears when the user clicks the ``Add Feedback'' button (\Cref{fig:improvementworkflow}E). Users can navigate to any previous revision point and create new branches as needed. For a more detailed screenshot of the revision tree, see \Cref{fig:revision} in \Cref{app:extended}.

\topic{Automated Operation Decomposition Assistance} After each pipeline run, \docwrangler automatically evaluates output accuracy using an LLM-as-judge approach in the background~\cite{zheng_judging_2023, shankar_who_2024}. \docwrangler samples five resulting documents and queries gpt-4o-mini in a single prompt to assess whether outputs meet criteria, returning \ttt{True} or \ttt{False}. For \ttt{False} results, \docwrangler then queries the LLM again to generate specific failure reasons and improvement suggestions, which are displayed as non-intrusive notifications. When users accept decomposition suggestions, \docwrangler invokes \docetl's accuracy optimizer to generate and test multiple candidate plans, or decomposed versions of the operation, and return the highest-accuracy plan (according to the LLM-as-judge). All optimization logs are streamed to the UI in real-time.

\topic{Managing Context Windows for AI-Assisted Features} Our AI-assisted features (refinement, decomposition, and the general-purpose chatbot) must handle limitations of the context window, or the maximum input size an LLM can process in one request. Since we include sample documents with every LLM interaction to provide necessary background, our messages often exceed the context window limit (128,000 tokens for gpt-4o-mini). To address this, we dynamically reduce content size before invoking the LLM. We first calculate the total token count of the entire conversation using \ttt{gpt-tokenizer}~\cite{gpt-tokenizer}. If it exceeds the limit, we determine how many tokens to remove and distribute this reduction equally across the sample documents in the first message only---maintaining subsequent conversation history intact. For each document, we remove text from the middle while preserving beginnings and endings, replacing the removed content with an ellipsis. Essentially, as the conversation history grows, documents progressively lose more middle content to accommodate new messages within the context window. We specifically preserve document beginnings and endings because introductions typically contain key metadata and conclusions often summarize content, both important for maintaining document context for the LLM.

\section{User Study}
\label{sec:method}

\begin{table}
\Description{Table showing participant demographics and their selected tasks. Columns include participant ID, background, dataset used, and tasks. Rows show participants from diverse fields like Data Science, AI, and Medicine. Each participant selected a dataset, such as Medical Transcripts or Privacy Policies, and one or more tasks to perform, labeled T1 through T4.}
\caption{Participant demographics and study tasks.}
\vspace{-5pt}
\label{table:demographics}
\footnotesize
\begin{tabular}{p{0.3cm}p{3.25cm}p{2.1cm}p{1cm}}
\toprule
\textbf{ID} & \textbf{Background} & \textbf{Dataset} & \textbf{Tasks} \\
\midrule
P1 & Data Science, AI & \task{D1Color}{Medical Transcripts} & T1, T4 \\
P2 & Data Visualization & \task{D1Color}{Medical Transcripts} & T2, T3 \\
P3 & Operations Management & \task{D1Color}{Medical Transcripts} & T1, T4 \\
P4 & Data Science & \task{D1Color}{Medical Transcripts} & T2, T3 \\
P5 & Machine Learning Engineering & \task{D3Color}{Safety Records} & T1–T3 \\
P6 & Data Science & \task{D2Color}{Pres. Debates} & T2, T4 \\
P7 & Data Science & \task{D1Color}{Medical Transcripts} & T2, T3 \\
P8 & Data Science & \task{D4Color}{Privacy Policies} & T1, T2 \\
P9 & Medicine & \task{D1Color}{Medical Transcripts} & T2, T3 \\
P10 & Data Science & \task{D2Color}{Pres. Debates} & T1, T3 \\
\bottomrule
\end{tabular}
\vspace{-5pt}
\end{table}

\begin{table}
\Description{Table listing four datasets and their associated semantic data processing tasks. The datasets are Medical Transcripts, Presidential Debates, Safety Records, and Privacy Policies. Each dataset has a list of tasks such as extracting patient demographics, identifying humorous quotes, or extracting GDPR mentions. Tasks are labeled T1 to T4 for each dataset.}
\caption{Datasets and tasks used in our study.}
\vspace{-5pt}
\label{table:tasks}
\footnotesize
\begin{tabular}{p{1.9cm}p{6.1cm}}
\toprule
\textbf{Dataset} & \textbf{Tasks} \\
\midrule
\task{D1Color}{Medical Transcripts} &
T1: Extract patient name, age, and gender \\
& T2: Analyze known risk factors for illnesses \\
& T3: Analyze symptoms and associated medical advice \\
& T4: Analyze patient discomfort vs. illness type \\
\midrule
\task{D2Color}{Presidential Debates} &
T1: Extract humorous quotes \\
& T2: Track how discussion of topics changes over time \\
& T3: Identify evaded questions/topics \\
& T4: Identify topics discussed by each party \\
\midrule
\task{D3Color}{Safety Records} &
T1: Extract incident location \\
& T2: Extract involved persons \\
& T3: Classify report behavior type \\
\midrule
\task{D4Color}{Privacy Policies} &
T1: Extract CCPR and GDPR mentions \\
& T2: Extract retention durations with cited regulations \\
\bottomrule
\end{tabular}
\vspace{-5pt}
\end{table}

Using \docwrangler as a design probe, we sought to understand both the tool's effectiveness and how people make progress in semantic data processing through a task-based, think-aloud study.

\topic{Participants and Recruitment} We recruited 10 participants via a call on the \docetl Discord server. While small, this size is corroborated by prior work suggesting that even five participants can uncover valuable usability insights~\cite{nielsen1994usability}.  All participants had prior experience using LLMs, with varied backgrounds in data processing. Four had previously used \docetl. Roles included software and ML engineers, data scientists, startup executives, medical professionals, and graduate students in CS and social science. All participants consented to audio and video recording. \Cref{table:demographics} summarizes their backgrounds and datasets.

\topic{Study Protocol} Following approval from our Institutional Review Board (IRB), we conducted one-hour long task-based sessions with each participant via video-conferencing over Zoom. Each session began with a 15-minute onboarding demo using a simple map-reduce pipeline. This walkthrough introduced \docwrangler prerequisites and features: dataset inspection, pipeline construction, pipeline execution, output inspection, and use of the AI assistant chatbot. Then, participants were given a choice between two datasets---medical transcripts and prosidential debates for the study tasks. Six participants selected the former, while two participants selected the latter. Two participants brought their own document datasets for the study; P4 brought a dataset on public safety records, while P8 brought a dataset with web-scraped privacy policies. Participants were then asked to complete at least two predefined tasks on their chosen dataset (\Cref{table:tasks}). They spent 40 or more minutes completing tasks, and were encouraged to think aloud and ask questions as needed. After the tasks, we collected feedback through open-ended reflections and Likert-scale ratings on IDE usability and comfort. We also asked follow-up questions about challenges with LLMs and expressing intent in \docwrangler. 

\topic{Analysis} We used Zoom’s auto-generated transcripts, supplemented by our notes, to document each session. Four authors independently conducted open coding of notes and transcripts, followed by two rounds of axial coding~\cite{taylor2015introduction} to identify recurring themes. 

\section{User Study Findings}
\label{sec:findings}

In this section, we discuss our qualitative findings from our study. Informally, we were interested in studying how participants design, iterate, evaluate, and debug their pipelines in \docwrangler to navigate the gulfs of comprehension, specification, and generalization in \Cref{fig:cycle}. All participants ($n=10$) found the IDE to be useful, and appreciated \docwrangler's potential for diverse document analysis tasks. {\bf \em Participants rated the ease of using \docwrangler for the semantic data processing task highly on a 7-point Likert scale (median = 6.5, mode = 7), with 80\% selecting 6 or 7 and no ratings below 5.} Our key findings are as follows:

\begin{itemize}[nosep, left=0pt]
\item Users manipulate semantic operations to aid validation, by requesting explanatory rationales or overly specific output attributes, and transforming open-ended tasks into structured classification problems;
\item Users opportunistically realign their pipelines by discovering both limitations of and possibilities with LLMs, pivoting between task refinement and goal reformulation; and
\item Users struggle with making sense of LLM-generated outputs at scale, requiring better provenance tracking and visualization tools tailored to semantic data processing.
\end{itemize}

\begin{figure}
\centering
\includegraphics[width=0.9\linewidth]{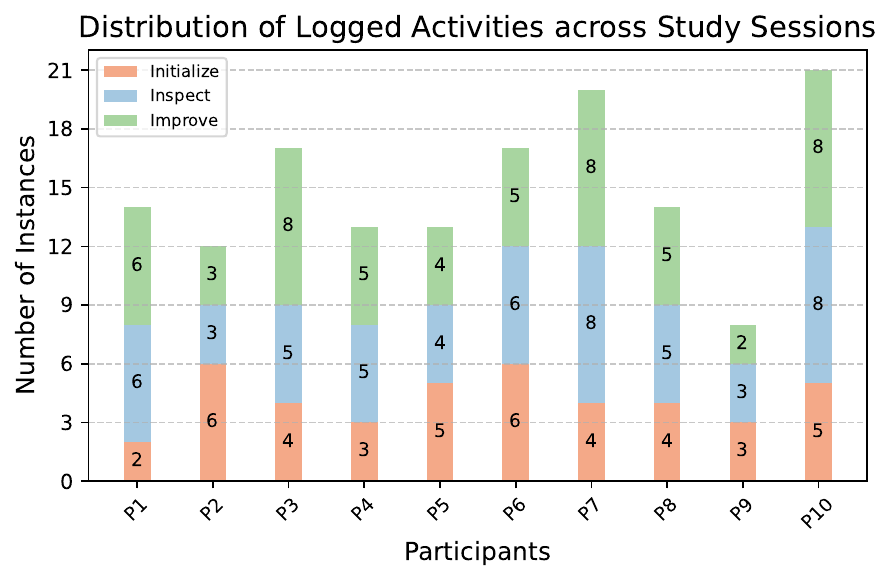}
\vspace{-10pt}
\Description{Bar chart titled ``Distribution of Logged Activities across Study Sessions.'' The x-axis lists participants P1 through P10, and the y-axis shows the number of instances, ranging from 0 to 21. Each participant has a stacked bar broken into three colored segments: orange for Initialize, blue for Inspect, and green for Improve. Most participants show a mix of all three activity types, with some bars reaching higher counts due to increased engagement in the Improve phase.}
\caption{Participant actions by phase: {\em Initialize} (create/edit operations), {\em Inspect} (review outputs), and {\em Improve} (refine prompts or decompose operations). Engagement was balanced across phases, with frequent transitions between them.}
\vspace{-10pt}
\label{fig:activity-split}
\end{figure}

\noindent We subsequently expand on our findings.

\subsection{Users Manipulate Semantic Operations to Bridge the Gulf of Comprehension}
\label{sec:findings-repurpose}

While participants differed in how they constructed their pipelines---most ($n=7$) built and inspected outputs of one operator at a time, while others ($n=3$) created complete \ttt{map}-\ttt{reduce} (i.e., extract-summarize) pipelines before inspecting outputs---participants similarly repurposed semantic operations in creative ways to make sense of outputs. We observed two main approaches: modifying LLM outputs to be easier to scan and interpret, and converting open-ended tasks into more structured classification problems. 

\subsubsection{Users modify output content and formatting to make it easier to interpret LLM behavior} 
\label{sec:findings-sample-and-structure}

To better understand LLM behavior at a glance, participants often adjusted operation outputs for interpretability. P1 and P8 added ``reasoning'' attributes to operation output schemas, with P1 noting this would \blockquote{force the LLM to explain its process.} P6 created separate \ttt{map} operations to generate summaries of extracted attributes, observing that these could \blockquote{reduce the amount of data that you're [manually] processing and validating in the pipeline, considerably.} Participants also adjusted the presentation of outputs to support manual validation. Some (P2, P3) used \workflow{\bf In-Situ User Notes} to request output reformatting (e.g., bulleted lists), making results easier to scan. P8 added boolean indicator attributes (e.g., \ttt{has\_GDPR\_mention}) to filter by specific mentions in the output table, then applied a \ttt{reduce} operation to summarize the behavior of the preceding \texttt{map}---without reviewing its full output manually. Importantly, these added attributes (e.g., rationales, summaries, indicators) were not used as final task outputs. Instead, they served to help participants verify whether the LLM had correctly interpreted their intent---bridging the specification gulf illustrated in \Cref{fig:cycle}. Reviewing these structured outputs often triggered \workflow{\bf In-Situ User Notes}, especially when outputs exposed interesting or ambiguous patterns. For instance, P9 (a doctor) noticed frequent mentions of ``back pain'' when operation outputs were formatted as bulleted lists. Drawing on their medical knowledge, they recognized this as a commonly reported symptom and filtered outputs containing ``back pain'' in the output table viewer to read mentions of ``back pain'' in the original documents and analyze its context with other reported conditions. They then annotated those outputs using \workflow{\bf In-Situ User Notes} to clarify context for the LLM; e.g., explaining the difference between ``acute'' and ``chronic'' back pain; the latter co-occurring with long-term conditions.

\subsubsection{Users transform open-ended tasks to classifier-like tasks} 
\label{sec:findings-map-as-classifer}

\begin{figure}
\centering
\includegraphics[width=0.8\linewidth]{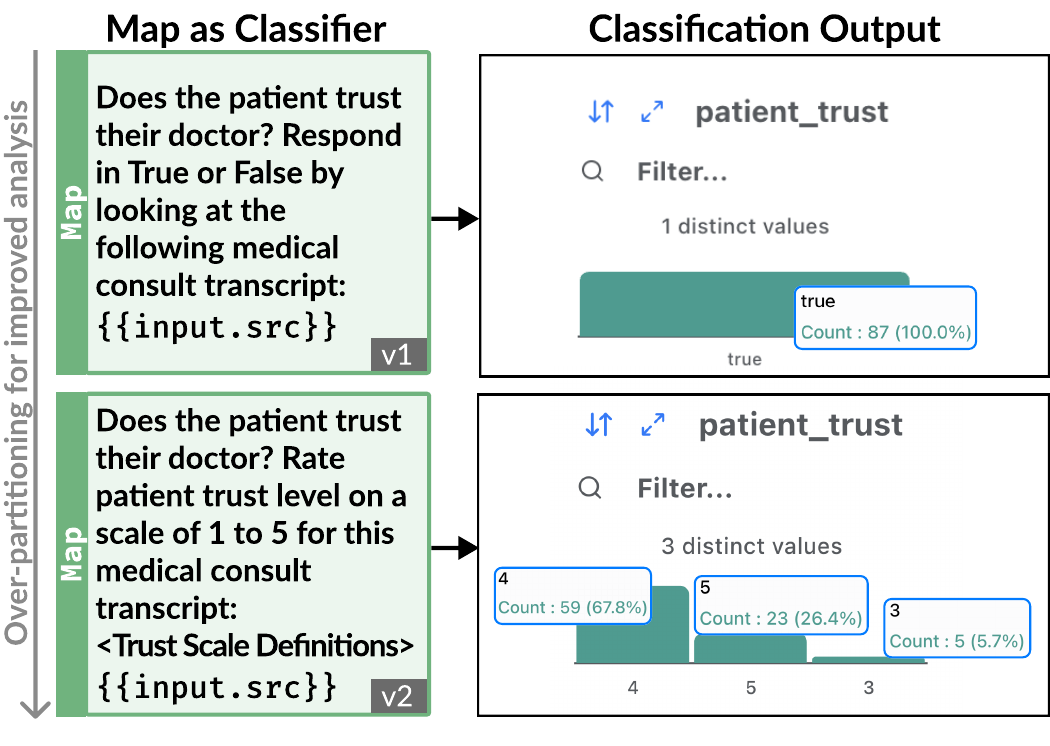}
\Description{The figure shows two side-by-side prompt-and-output pairs labeled v1 and v2. On the left, the prompts are shown inside green boxes under the heading ``Map as Classifier.'' v1 asks if the patient trusts their doctor with a True/False response. v2 asks for a trust rating from 1 to 5. On the right, under ``Classification Output,'' v1 shows a single bar labeled ``true'' with 87 responses. v2 shows a bar chart with five values from 1 to 5, with counts distributed across the scale. An arrow and label indicate a shift from binary to scaled classification for improved analysis.}
\caption{Participants converted open-ended prompts into classification tasks for easier debugging. Here, P1 used a binary version (v1) that yielded uninformative results (100\% ``true''), then switched to a 5-point scale (v2) for higher accuracy. P1 could then apply a code-based rule (e.g., score \textgreater4) to interpret the results as a binary outcome.}
\label{fig:classification-churn}
\vspace{-10pt}
\end{figure}

To make validation more manageable, participants often reframed open-ended tasks as classification problems (P1, P4, P5, P8, P9, P10). P1 described this as \blockquote{treating the LM like a classifier}---examining class-based outputs to check for meaningful differences. For example, when analyzing doctor-patient trust, P1 initially used a free-form \ttt{trust\_summary} attribute, but added a boolean \ttt{trust} attribute to validate results more easily via a histogram. As shown in \Cref{fig:classification-churn}, the LLM labeled all examples as ``true,'' so P1 switched to a 5-point Likert scale for more granularity. While the scores were still skewed high, the distribution revealed more variation. P1 noted that they could apply a code-based rule (e.g., score~\textgreater4) to interpret the results as binary, and used the differences between low and high scores to identify behavioral signals (e.g., patient stuttering), which they annotated using \workflow{\bf In-Situ User Notes}.

Even when tasks didn’t lend themselves to ``validation by histogram,'' categorical attributes enabled more systematic inspection.  For example, P10 wrote an operation to extract a list of quotes containing logical fallacies from each debate transcript, then added a \ttt{fallacy\_type} attribute to label each quote (e.g., strawman, ad hominem). Subsequent grouping by type made it easier for P10 to spot errors, since the LLM performed better on some fallacies than others. Some participants developed more sophisticated approaches; e.g., P5 requested the AI chatbot assistant to automatically generate a taxonomy of document types to use in a \ttt{map} operation, and P9 created a hierarchical taxonomy themselves and edited their prompt to include this taxonomy.

\subsection{Users Iteratively Refine Pipelines to Navigate All Gulfs}
\label{findings-opportunistic-realignment}

Unlike typical data science workflows where users begin with exploratory data analysis~\cite{muller2019data}, all participants skipped manual document review and jumped straight into writing \ttt{map} operations. As they inspected outputs, they frequently revised their pipelines in response to what they saw---what we call {\em opportunistic realignment}. This realignment helped users bridge gulfs in \Cref{fig:cycle}---the specification gulf, by refining prompts and operations to better express their intent; and the generalization gulf, by tuning their workflows to improve alignment with the quirks of their specific data.

Realignment occurred when users discovered either limitations in what the LLM could do, prompting debugging or operation decomposition, or possibilities that surfaced through surprisingly useful outputs. In both cases, users adjusted their goals or prompts based on what they learned from the system’s behavior. Interestingly, even though the \workflow{\bf Operation Decomposition} feature was designed to help address the generalization gulf, users sometimes adopted it as a way to improve specification too---using suggestions to rethink how they framed their tasks or restructure their prompts.

\subsubsection{LLM limitations lead users to reframe their goals}
\label{sec:findings-discovering-limitations}

Participants often encountered LLM limitations that forced them to pursue alternate ways of accomplishing the same high-level task (P1, P4, P6, P9). These alternate paths often emerged after inspecting outputs and realizing the LLM misunderstood the original intent. For instance, in a medical information extraction task, P4 wanted to extract unique symptoms that patients mentioned in the conversation transcripts. However, the LLM frequently returned near-duplicate entries like ``pain when pressure applied,'' ''pain when lying down,'' and ``pain when sleeping.'' P4 instead had wanted a small set of unique symptoms, grouping related variants in parentheses, e.g., ``pain (when pressure applied, when lying down)''. P6 faced a similar issue in a political debate analysis task. Their initial prompt produced long-winded topic summaries that buried the main points. They hadn’t originally planned to constrain output length, but after reviewing the verbose outputs, realized they preferred more concise descriptions---and added explicit brevity instructions.

In realigning with the LLM’s interpretation of the pipeline, users discovered limitations in common prompt engineering strategies (P1, P4). P1 initially wanted to include successful output examples (and corresponding documents) in their prompts~\cite{wang2020generalizing} but realized that the documents contained too much irrelevant information, making it difficult for the LLM to infer what made these outputs successful. P1 spent considerable time reasoning about the LLM's inferential capabilities, repeatedly shifting between output checking, providing notes, and prompt refinement---disrupting flow~\cite{tankelevitch2024metacognitive}.

Some participants redistributed work between semantic (LLM) and code-based components (P2, P4). When analyzing doctor-patient transcripts, P2 and P4 initially used a semantic \ttt{map}-\ttt{reduce} pipeline to extract and summarize symptoms. However, the \ttt{reduce} operation often overgeneralized, collapsing distinct mentions like knee pain'' and back pain'' into simply ``pain.'' To preserve symptom specificity, they replaced the semantic \ttt{reduce} with a code-based \ttt{reduce} that used string matching to tally each extracted symptom (maintaining distinctions between similar symptoms like ``knee pain'' and ``pain in the knee'' that would be counted as separate items even though they represent the same underlying condition). P4 then added a second \ttt{map} operation that used an LLM to generate a comprehensive report incorporating both the raw symptom extractions and their frequency counts---giving the LLM the discretion to merge related symptoms while using the frequency data to highlight which symptoms were most common across the dataset. 

Over time, participants effectively navigated the specification gulf (\Cref{fig:cycle}) through iteration. P9 reflected that their process was surprisingly systematic---\blockquote{more like iteration and problem articulation [based on what the LLM can do], not trying random things.} In this way, LLM interaction became a medium for clarifying analytical goals---echoing Schön's notion of reflection-in-action~\cite{schon1979reflective}.

\subsubsection{Unexpected possibilities can also shift analytical goals}
\label{sec:findings-discovering-possibilities}

Not all pivots were due to limitations. Some users shifted direction after spotting surprising or useful patterns in the LLM's outputs. These ``serendipitous'' findings weren’t requested explicitly, but appeared occasionally, revealing new opportunities for analysis. For example, P9 initially wanted to extract symptoms from doctor-patient transcripts, but noticed that the LLM also surfaced relevant patient history for a few outputs. They found this helpful---it both improved LLM accuracy and made outputs easier to validate---so they updated the prompt to also extract medical history. Others (P6, P10) discovered desirable output structures they hadn't asked for---like bulleted lists or Markdown formatting---and revised their prompts to consistently request those formats.

Participants also encountered meaningful patterns or categories in the data that they hadn't explicitly asked for (P1, P5, P7-P10). For instance, while analyzing public safety records, P5 noticed that some outputs described the roles of people involved in each incident. They realized these roles could reveal the type of document---such as a witness statement versus a legal document---and used this insight to refine their pipeline. P9 became aware of this emerging process and coined the term {\bf \em ``prompt rubber ducking''} to describe how interacting with LLMs helped them figure out what questions to ask about their data. In this way, semantic data processing pipelines don’t just answer predefined questions, they also help shape users' understanding of {\em what} questions are worth asking---perhaps similar to the ``berry picking'' model of information seeking, where users iteratively refine their search as they gain new insights~\cite{bates1989design}.  

Despite frequent shifts in focus (e.g., inspecting, reflecting, refining), users maintained a strong sense of progress throughout. This stemmed from two factors. First, system responsiveness allowed rapid iteration, impressing users (P1, P3-P5, P8, P9). For example, P8 expanded a GDPR compliance analysis to include CCPA patterns simply by changing ``GDPR'' to ``CCPA'' in the prompt, and results appeared within seconds. Second, output schemas acted as what P8 described as \blockquote{speed breaks}, slowing exploration just enough for meaningful reflection. We observed that several participants (P2, P6, P7) visibly slowed down when writing these schemas. Output schemas served as ``semantic'' type checks---a form of validation analogous to type checks in traditional programming~\cite{coblenz2021plushci}.

\subsubsection{Decomposition Actually Supports Both Generalization and Comprehension} \label{sec:findings-decomposition}

Although the \workflow{\bf Operation Decomposition} feature was originally designed to address the generalization gulf in \Cref{fig:cycle}---by helping users improve output accuracy across diverse documents---participants also used it to navigate the comprehension gulf, helping them understand how the LLM interpreted their data. The decomposition analysis inspired alternate task explorations, temporarily ``unblocking'' the user (P2, P4, P5, P9). All participants reviewed decomposition suggestions at least once, but only some (P2–P4, P8, P10) allowed \docwrangler to automatically apply them. A key factor in whether participants accepted suggestions was their confidence in implementing the changes on their own. When users felt capable of restructuring an operation, they preferred to use decomposition insights as inspiration---either to manually rewrite prompts or provide \workflow{\bf In-Situ User Notes} (P2, P3). But when the suggestions felt too complex or outside their expertise, they opted into automatic decomposition (e.g., P5 for entity resolution, or P4 when they \blockquote{wanted the system to do it better}). P4 explicitly invoked decomposition when they could not articulate the issue as a pipeline refinement, explaining \blockquote{I feel like there's a gap in me understanding how to reconcile what's being suggested and what's being set up.}

Importantly, even when participants chose to implement changes themselves rather than accepting automatic decomposition, they still benefited from \docwrangler's ability to identify issues they might have missed. For example, while manually verifying every extraction across all documents was impractical, it was much easier for participants confirm whether the decomposer's specific notifications about missed information were accurate (P4, P9). AI-assisted analysis, in this way, helps users bridge the comprehension gulf by pinpointing specific data they might otherwise overlook.

\subsection{Participants Identified Gaps in Tooling}
\label{sec:findings-tools}

Participants pointed out specific limitations in \docwrangler that made it harder to interpret or reason about LLM outputs. These gaps often came up as feature requests, highlighting needs for better support with pattern discovery, provenance tracing, and operation-specific visualization.

\topic{The Necessity of Bottom-up Validation} All participants recognized that they were spending a lot of time validating outputs, and some explicitly requested to automate this with LLM-as-a-judge approaches (P4, P7). Rather than defining evaluation criteria upfront or using a top-down approach, as prior work suggests~\cite{shankar_who_2024, kim2024evallm, ma2024you}, participants (P2, P4) preferred to discover criteria bottom-up through hands-on review---similar to qualitative coding~\cite{taylor2015introduction}, but prompted by LLMs identifying both interesting patterns and outliers. P4 considered writing an operation to validate the previous operation and began thinking of a rubric to put in the prompt, but realized they’d need to review many outputs first to come up with this rubric, and then repeat this process after each pipeline change. With goals shifting frequently, this felt impractical. Others (P1, P2) requested tools for organic pattern discovery, such as P1’s request for side-by-side comparisons of LLM outputs across diverse, automatically-selected documents, to more easily spot issues and provide in-situ notes.

\topic{Additional Support for Inspecting Provenance} Participants wanted easier ways to trace how LLM outputs were derived from the source text without having to manually search documents or write custom checks (e.g., a code-based \ttt{filter} after each \ttt{map}). Some (P2, P8, P9) requested that outputs be directly linked to their source text, with the relevant span highlighted. However, simple checks for the presence of terms don't guarantee the operation was performed correctly. For example, the LLM listed ``muscle aches'' as a symptom, even though it appeared only in the doctor’s question---not as something the patient reported, showing how surface-level matches can be misleading (P9). Moreover, users wanted \docwrangler to trace {\em errors} to their source. For example, when P4 reviewed extracted medication information, there were incorrect dosages in the source text (e.g., ``200g'' instead of ``200mg'') that the LLM faithfully reproduced. This highlights how LLM pipelines blur the boundary between data cleaning and analysis, unlike traditional workflows where these phases are less intertwined~\cite{muller2019data, kandel2012enterprise}. We never observed participants creating dedicated cleaning operations---perhaps, as P2 mentioned, because they expected LLMs to implicitly clean data (e.g., recognize that 200g'' is not a valid medication dose).

\topic{Operation-Specific Visualization Tools} Users requested richer visualizations beyond basic histograms and bar charts (P4, P5, P7, P10), with needs varying by operation type and data domain. P7 requested LLM-generated custom charts. P4 asked for \docwrangler to track output distribution {\em changes} between iterations, particularly when using map operations as classifiers (\Cref{sec:findings-map-as-classifer}).

\section{Real-World Deployment and Usage}
\label{sec:deployment}

To complement our qualitative user study, we deployed \docwrangler as a public web application, collecting telemetry data from over 1,500 pipeline executions across two months. We used \docwrangler itself to analyze this telemetry data, manually verifying a sample of 50 extractions and classifications.\footnote{Our analysis involved three \texttt{map}-\texttt{reduce} pipelines with gpt-4o; one for each topic (e.g., usage domains and document types). Our total analysis cost was \$9.39 USD.} We will discuss the domains and document types in which \docwrangler was used, common task patterns and pipeline structures, how pipelines evolved, and how users engaged with AI assistance features. 

\topic{Pipeline Data, Task, and Model Patterns} \docwrangler was used across a range of professional domains. Common applications included legal (e.g., contract clause extraction), healthcare (e.g., analyzing clinical case studies), finance (e.g., parsing invoices and budgets), and education (e.g., generating test questions from textbooks). Other frequent use cases involved customer feedback analysis, government document processing, and media or news content analysis. About 50\% of documents were semi-structured with hierarchical sections, though often inconsistently formatted. Unstructured text was the next most common, typically in PDF format. Pipelines spanned 9 languages: English, Spanish, Chinese, German, Russian, Japanese, Italian, Greek, and Persian.

Most pipelines centered on a few key task types. Extraction was most common, appearing in over half of all pipelines---from pulling out existing structure to identifying semantically meaningful entities in the absence of structure. Classification ranked second, typically categorizing documents or content. Summarization usually followed these steps using reduce operations. Most workflows followed a simple pattern: extract, classify, then summarize, but some domains exhibited different patterns. For example, in business analysis, users built complex pipelines (sometimes with 15+ operations) that performed sequential extractions---first identifying entities like departments and processes, then mapping relationships, and finally generating reports. In the legal domain, pipelines were often multi-step as well, but focused on extracting varied or less-related entities, such as claims and case law references. For research domains, pipelines often involved more open-ended pattern discovery across documents with \ttt{map} operations---then more structured  \ttt{map}-\ttt{reduce} pipelines. Most pipelines were simple: 75\% had two or fewer operations, and 90\% had no more than three. But 5\% had more than five operations, and a few pipelines exceeded 30, reflecting highly customized workflows.

Pipelines used different AI models. gpt-4o-mini (our default model) was the most common model, used in 82\% of pipelines.  gpt-4o appeared in 12\% of pipelines, typically for more complex workflows or in combination with gpt-4o-mini. Gemini models were used in 17\%, often for PDF processing. Claude-3.5-sonnet showed up in 9\%, mainly for classification. 18\% of pipelines used multiple models.

\topic{Pipeline Evolution Patterns} We analyzed how users modified their pipelines over time by examining consecutive pipeline versions created within 5-minute intervals. We observed three main evolution patterns: 53\% of pipelines grew more complex by adding operations or upgrading models; 18\% actually became simpler through operation consolidation or reduced sample sizes; and 29\% maintained the same operations while only changing prompts or output schemas. We also tracked how prompts evolved: 47\% became more specific with structured schemas and domain-specific language; 16\% moved toward simpler prompts (following the ``prompt rubber ducking'' strategy mentioned in \Cref{sec:findings-discovering-possibilities}); and 37\% maintained similar detail levels with minor refinements.

\topic{AI Assistance and Challenges} We tracked AI assistance features: prompt improvement (150 instances) and general AI chat (95 instances). Surprisingly, for prompt refinement, over half of these cases involved users refining their prompts without providing any notes. Users were invoking prompt refinement proactively to transform their vague, poorly-specified instructions into more concrete and executable prompts; for example, one user invoked prompt refinement to transform a prompt like \blockquote{Extract the information from the ledger} into a detailed 10-line specification identifying precise fields to extract and noting important context, such as \blockquote{these are scans of handwritten documents.} This reveals how users often struggle with formulating effective prompts from the outset. Through the AI chat, users asked general questions about syntax, workflow creation, and supported file formats. Some users still expressed frustration when their prompts didn't work as expected, with notes like \blockquote{MAP THEM ALL SYSTEMATICALLY!!!!!!!!} and \blockquote{WHY IS THIS SO HARD?!} These challenges suggest we need better guidance and clearer explanations of \docwrangler's capabilities and limitations, particularly for new users.

\section{Discussion}
\label{sec:discussion}

Here, we discuss what our findings mean for current data processing systems and human-AI collaboration more broadly.

\topic{Designing Better Semantic Data Processing Tools} \docwrangler is an early attempt at building an IDE for semantic data processing, which taught us a lot about how users interact with data analysis workflows with LLM-powered operators. Based on our findings, we reflect on opportunities for future semantic data processing systems.

When building \docwrangler, we realized that bridging the specification gulf involves two different challenges: finding the ``right'' question to ask (i.e., intent discovery) and clearly expressing that question in the pipeline (i.e., refining ambiguity).  While we tried to address both with the same features in \docwrangler, they are fundamentally different. Discovering intent is exploratory: users need to understand what's interesting or important in their data before formulating precise questions. For structured data, we already have effective tools for kick-starting exploration, by automatically discovering and surfacing typical patterns or outliers~\cite{siddiqui2016effortless, lee2021lux, wongsuphasawat2015voyager} or by letting users search for specific patterns through sketching or demonstration~\cite{battle2016dynamic, hochheiser2004dynamic, siddiqui2018shapesearch}. However, these approaches need reimagining for unstructured data and semantic processing. It's not clear what constitutes a ``document anomaly,'' or how to tell whether documents are meaningfully different in size, structure,  or content. Future systems could address this challenge and help proactively highlight patterns in the underlying documents, ultimately making it easier for humans to decide what they want to query. In contrast, clarifying intent---once identified---is about precise specification and addressing ambiguity, and may be more amenable to automation. Systems can suggest more precise wording and recommend better alignment with how LLMs interpret instructions (similar to how we supported prompt refinement in \docwrangler).

Additionally, our study revealed several debugging strategies that future systems could partially automate. For example, users frequently transformed unstructured extraction tasks into more structured ones to make validation easier (\Cref{sec:findings-map-as-classifer})---so, systems could detect when outputs lack structure and automatically suggest complementary structured attributes to aid validation. Similarly, when systems detect skewed output distributions (like P1's all-``true'' trust classifications in \Cref{fig:classification-churn}), they could automatically rewrite prompts to yield more useful variations.  \techreport{Systems could also automatically generate rationales as interpretability tools (\Cref{sec:findings-sample-and-structure}), by providing explanations, clustering or finding anomalous examples, and presenting these for review to help users spot intent-interpretation mismatches.}

Overall, semantic data processing, and the advent of LLMs, has the potential to change how we build the next generation of data systems~\cite{fernandez2023large} that combine both structured
and unstructured data processing capabilities. Our systems are no longer ``passive'' executors of fixed specifications like traditional data processing tools; instead, they must help users express and refine semantic needs through their pipelines~\cite{zeighamillm}. To make semantic data processing truly usable, we must design systems that actively bridge all three gulfs---between users' intentions, pipeline specifications, and underlying data---be it in a structured or unstructured context.

\topic{Semantic Data Processing as a Lens on Human-AI Collaboration} Semantic data processing serves as a rich domain to study effective human-AI collaboration. 
The interaction triangle shown in \Cref{fig:cycle} may generalize to other human-AI systems if we recognize that ``data'' might be much smaller or take different forms than document collections (e.g., a single essay, a piece of software). Consequently, our findings might offer insights for designing effective human-AI systems across domains.

First, we reflect on {\em how} to design AI systems. We observed users creating what creativity support tool (CST) research calls {\em epistemic artifacts}~\cite{tricaud2023revisiting}---exploratory objects that help users discover possibilities. Early pipeline iterations served as epistemic artifacts, helping users learn about their data rather than being final solutions, similar to how artists explore materials before creating finished works. Though \docwrangler wasn't intended to be a CST, it functioned as one because tasks had inherent fuzziness, and users learned about either their preferences or the task itself through LLMs. In this way, {\em any} system addressing ambiguous tasks may benefit from CST design principles; e.g., supporting exploration without predefined goals and allowing movement between different levels of abstraction~\cite{tricaud2023revisiting, li2023beyond}.  

Then, like CSTs, AI systems must preserve user agency~\cite{horvitz1999principles}.\footnote{We acknowledge that ``agency'' has been critiqued as an overly broad term in HCI~\cite{bennett2023does}; here we also use it in this general sense, rather than referring to a specific definition.} To maintain this agency, prior work suggests using shared representations of information that both humans and AI can modify~\cite{heer2019agency}, or creating different representations that give users varying levels of control~\cite{Satyanarayan2024Intelligence}. Our work adds another perspective: creating moments for user reflection~\cite{schon1979reflective}. For example, our in-situ notes feature allowed (perhaps even {\em encouraged}, due to its ubiquitous presence) users to record thoughts. We also unintentionally created reflective spaces: e.g., output schema definitions slowed users down, prompting their reflection on their goals (P8).

We also reflect on {\em what} to design for AI systems. Each gulf in \Cref{fig:cycle} represents a distinct design challenge, with multiple possible approaches to address each one (not just what \docwrangler proposes). Moreover, applications should offer features to reduce the burden of navigating all gulfs, but specialized domains may benefit from specialized approaches. For instance, AI image generators already bridge the specification gulf with controls for attributes like ``realism,'' ``style,'' and ``composition''\cite{midjourney, brade2023promptify, rombach2022high}, but these systems can also address the generalization gulf by automatically flagging semantic inconsistencies like anatomical errors or impossible scenes---constraints users shouldn't need to explicitly state if they want high realism. In another example, an AI coding assistant could use compilation errors to ``self-refine'' its generated code~\cite{madaan2023self} (bridging the generalization gulf), while also maintaining these errors as notes and summarizing them (aiding the comprehension gulf).  Or, consider an AI vacation planner: interactive visualizations of possible trips could help users comprehend both the recommendations and what the AI has inferred about their preferences.

Overall, based on our work, we present the following recommendations for building effective human-AI collaborative systems: {\em (i)} design for epistemic artifacts that help users explore and understand their problem space; {\em (ii)} create intentional moments for reflection that maintain user agency amidst rapid AI-driven iterations; and {\em (iii)} develop features that explicitly address each of the three gulfs---specification, generalization, and comprehension---recognizing that bridging one gulf can indirectly help with others. 

\topic{Limitations} We reflect on some limitations of our work. First, our participants were all tech-savvy and had used LLM tools before, with most having programming experience (though three had not heard of map-reduce before). This may have affected how easily they adapted to our system. Future work should include users with varying technical backgrounds. Second, we only observed sessions lasting 1-2 hours, while real document analysis often spans days or weeks. Longitudinal studies could reveal how behavior evolves over longer periods and in team settings. In our online deployment, as users needed to provide their own API keys for LLMs, our user base is self-selecting---primarily comprising individuals who discovered \docwrangler through blog posts and technical talks. Additionally, some users, particularly those with more experience, run \docwrangler on their own infrastructure, for which we lack telemetry data. Finally, our findings come from observing users with our specific system. While we believe the patterns we identified reflect fundamental aspects of human-AI interaction, more research is needed to understand how they manifest across different interfaces, LLM capabilities, and domains.

\section{Conclusion}
\label{sec:conclusion}

This paper introduced \docwrangler, a mixed-initiative IDE for semantic data processing, where LLMs power familiar data operations like \ttt{map} and \ttt{reduce} on unstructured text. We contributed three novel features to bridge the gulfs between users, their data, and their semantic data pipelines: in-situ user notes, LLM-assisted prompt refinement, and LLM-assisted operation decomposition. Our evaluation through both a qualitative user study and large-scale online deployment provides insights into how people adapt to and use \docwrangler for their tasks; particularly, how they write pipelines for the purpose of learning more about their data or LLM capabilities. Finally, our work provides a reflection on how data systems could be redesigned for the LLM era and what semantic data processing can teach us about effective human-AI collaboration more broadly.

\begin{acks}
We are grateful to Shm Almeda, Ian Arawjo, Timothy Aveni, Quentin Romero Lauro, Yiming Lin, HC Moore, James Smith, J.D. Zamfirescu-Pereira, and Sepanta Zeighami for their thoughtful feedback and valuable discussions on our findings. We thank Preetum Nakkiran for inspiring the islands concept in Figure 1 and providing moral support during its creation. Special thanks to our study participants, whose insights and engagement were instrumental in understanding how users interact with semantic data processing systems. This work was supported by the National Science Foundation (grants DGE-2243822, IIS-2129008, IIS-1940759, IIS-1940757, 1845638, 1740305, 2008295, 2106197, 2103794, 2312991), funds from the State of California, funds from the Alfred P. Sloan Foundation, and EPIC lab sponsors (G-Research, Adobe, Microsoft, Google, Sigma Computing). Additional support was provided by Amazon and CAIT. We thank Modal Labs for their generous compute credits. SS is supported by a National Defense Science and Engineering Graduate (NDSEG) Fellowship.
\end{acks}
\bibliographystyle{ACM-Reference-Format}
\bibliography{references}

\clearpage
\appendix
\section{Extended Feature Walk-Through}
\label{app:extended}

\begin{figure}
    \centering
    \includegraphics[width=0.9\linewidth]{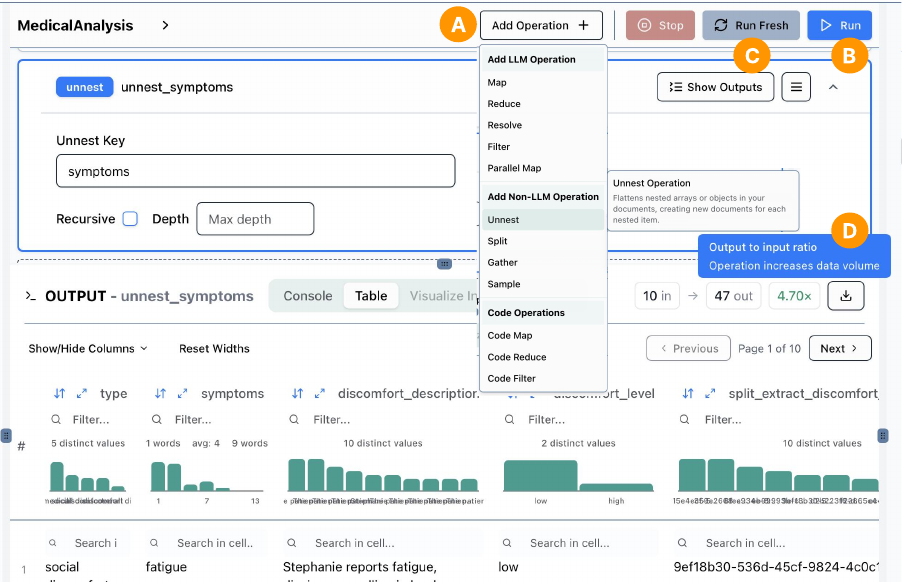}
    \Description{The figure shows four labeled interface elements, A to D, depicting the process of adding an unnest operation. A highlights a dropdown where the user selects ``Add Operation'' and chooses ``unnest.'' B and C show two buttons: ``Run,'' which uses cached outputs, and ``Run Fresh,'' which reprocesses everything from scratch. D shows the output table after applying the unnest operation, with symptoms displayed as separate rows and an increase in row count from 10 to 47. Bar charts and metadata are visible in the output viewer.}
    \caption{Adding an \ttt{unnest} operation to analyze symptoms individually. (A) User adds a new operation. (B) ``Run'' button executes with previously-cached operation outputs, while (C) ``Run Fresh'' reprocesses all operations from scratch. (D) Output viewer shows operation selectivity (10 documents expanded to 47; a $4.7\times$ increase), with each symptom as a separate row preserving associated metadata.}
    \label{fig:unnest}
\end{figure}

\begin{figure}
    \centering
    \includegraphics[width=0.9\linewidth]{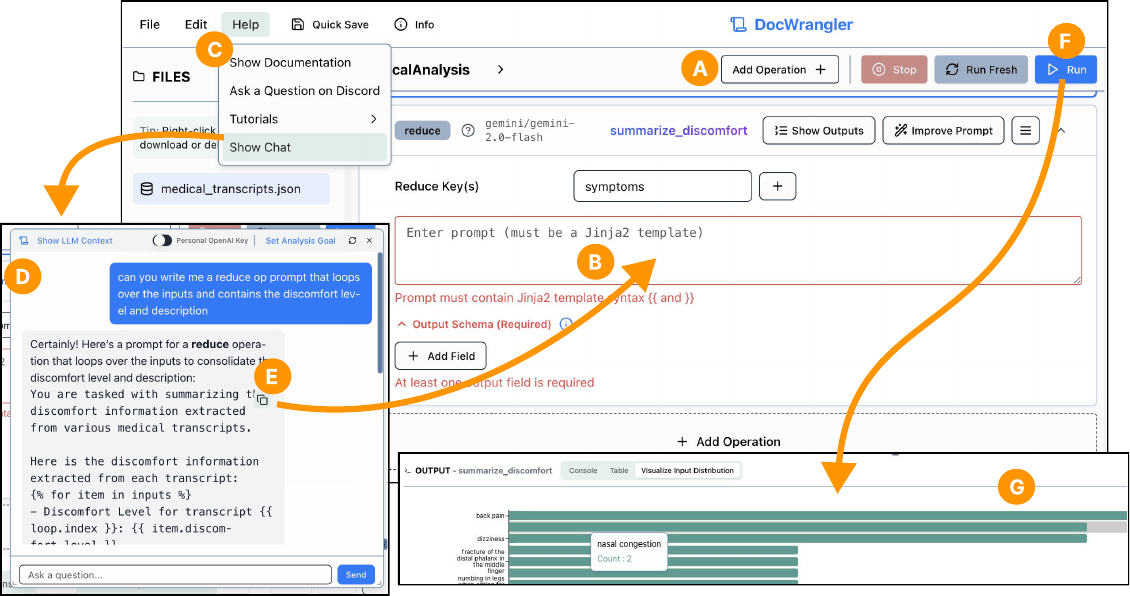}
    \Description{The figure includes three interface screenshots labeled with seven steps, A to G, showing how a reduce operation is added. In the top-left screenshot, A shows the user opening the ``Add Operation'' menu, and B highlights a form where the user enters the group-by field. In the lower-left screenshot, C and D show the help panel with the AI assistant, and E highlights the user copying a suggested Jinja expression. The large screenshot on the right includes F, where the user clicks ``Run,'' and G, which displays a bar chart visualizing the summarized symptom data. Orange arrows connect the steps across screenshots to show the full workflow.}
    \caption{Adding a \ttt{reduce} operation to summarize symptom data. (A) The user adds a new operation. (B) They specify which field to group by. (C) The help menu shows the AI assistant. (D) The AI assistant helps with Jinja syntax. (E) User copies the assistant's suggestion to copy into the pipeline editor. (F) The user runs the pipeline. (G) The output includes a visualization of the distribution of symptoms.}
    \label{fig:reduce}
\end{figure}

Here, we continue the feature walk-through from \Cref{sec:interface-example-usage}, illustrating use of different operators and the AI assistant chatbot.

Recall that at the end of \Cref{sec:interface-example-usage}, the analyst has decomposed their first operation into the pipeline described in \Cref{fig:decompositionworkflow}. With the decomposed pipeline now reliable, the analyst resumes analyzing discomfort by symptom. They click ``Add Operation'' (\Cref{fig:unnest}A) and select an \ttt{unnest} operation, specifying ``symptoms'' as the attribute to flatten.\footnote{\texttt{unnest} is akin to the ``explode'' operator in Pandas.} After configuration, they click ``Run'' (\Cref{fig:unnest}B), which uses cached outputs from previous operations, though ``Run Fresh'' (\Cref{fig:unnest}C) remains available for complete reprocessing. The operation executes instantly, with the output viewer showing {\em ``10 in $\rightarrow$ 47 out''} and {\em ``$4.70\times$''} (\Cref{fig:unnest}D). Each row now represents a single symptom, with all other data attributes (the original transcript text and LLM-extracted attributes like discomfort level) copied from the source document that contained that symptom.

To summarize discomfort patterns by symptom, the analyst adds a \ttt{reduce} operation (\Cref{fig:reduce}A-B), specifying ``symptoms'' as the group-by attribute. Unsure about the how to write a Jinja template to loop over a group of documents, they access the chat-based AI assistant (\Cref{fig:reduce}C-D) and request help writing their prompt. Using the AI's suggested syntax (\Cref{fig:reduce}E), they run the pipeline (\Cref{fig:reduce}F) and examine the symptom distribution (\Cref{fig:reduce}G). They can switch to the table view for detailed LLM output inspection as needed. Inspection---and the data analysis, in general---proceeds for as long as the user would like.

\begin{figure}
    \centering
    \includegraphics[width=0.9\linewidth]{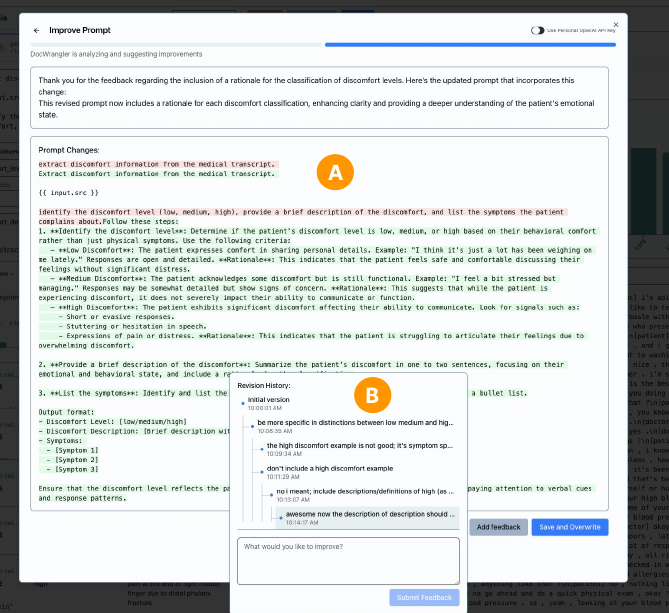}
    \Description{The figure contains a single screenshot of the prompt refinement interface with two labeled regions, A and B. Region A highlights a diff view showing changes between the original and revised prompt, with text edits color-coded for clarity. Region B shows a popup dialog displaying an interactive revision history tree, where users can view past versions and create new branches. The interface includes buttons to accept edits or continue revising.}
    \caption{\workflow{\bf Prompt Refinement} interface. (A) The interface visualizes the diff between prompt versions. (B) An interactive revision history tree allows users to view and branch from previous prompt versions.}
    \label{fig:revision}
\end{figure}

\end{document}